\documentclass{emulateapj}
\usepackage{mathptmx,subfigure,graphicx}
\begin{document}

\title{The lensed arc production efficiency of galaxy clusters: A
  comparison of matched observed and simulated samples}
\author{
Assaf~Horesh\altaffilmark{1}, 
Eran~O.~Ofek\altaffilmark{1}, 
Dan~Maoz\altaffilmark{1}, 
Matthias Bartelmann\altaffilmark{2}, 
Massimo~Meneghetti\altaffilmark{2}, 
Hans-Walter~Rix\altaffilmark{3}
}

\altaffiltext{1}
{School of Physics and Astronomy and Wise Observatory, Tel Aviv University, Tel Aviv 69978, Israel.}

\altaffiltext{2} {ITA, Universit\"at Heidelberg, Albert-Ueberle Str.
  2, D--69120 Heidelberg, Germany.}

\altaffiltext{3}
{Max Planck Institute f\"{u}r Astronomie, Koenigstuhl 17, D-69117 Heidelberg, Germany.}

\begin{abstract}
  
  We compare the statistical properties of giant gravitationally
  lensed arcs produced in matched simulated and observed cluster
  samples. The observed sample consists of 10 X-ray selected clusters
  at redshifts $z_{c}\sim 0.2$ imaged with HST by Smith et al.. The
  simulated dataset is produced by lensing the Hubble Deep Field,
  which serves as a background source image, with 150 realizations
  (different projections and shifts) of five simulated $z_{c}=0.2$
  clusters from a $\Lambda{\rm CDM}$ N-body simulation. The real and
  simulated clusters have similar masses, the real photometric
  redshift is used for each background source, and all the
  observational effects influencing arc detection in the real dataset,
  including light from cluster galaxies, are simulated in the
  artificial dataset. We develop, and apply to both datasets, an
  objective automatic arc-finding algorithm. We find consistent arc
  statistics in the real and in the simulated sample, with an average
  of $\sim 1$ detected giant $({\rm length~to~width~ratio} ~l/w\geq
  10)$ arc per cluster and $\sim 0.2$ giant luminous
  $(R_{ST}<22.3~{\rm mag})$ arc per cluster. Thus, taking into account
  a realistic source population and observational effects, the
  clusters predicted by $\Lambda$CDM have the same arc-production
  efficiency as the observed clusters. If, as suggested by other
  studies, there is a discrepancy between the predicted and the
  observed total number of arcs on the sky, it must be the result of
  differences between the redshift dependent cluster mass functions,
  and not due to differences in the lensing efficiency of the most
  massive clusters.
\end{abstract}


\section{Introduction}
\label{Introduction}

The statistical properties of giant gravitationally lensed arcs in
galaxy clusters are sensitive to clusters properties, such as mass
function, mass profile, and concentration, which in turn depend on
cosmology.  Cosmological models can therefore be tested, in principle,
by comparing the predicted numbers and properties of giant arcs with
observations.

Bartelmann et al. (1998; hereafter B98) compared the observed number
of giant arcs with predictions based on calculations of ray tracing
through galaxy clusters formed in N-body simulations, for various
cosmological models. Artificial sources were all placed at redshift
$z_{s}=1$, which is close to the average redshift of observed arcs
(Kneib \& Soucail 1996). B98 argued that using a single source
redshift is acceptable because the change in the critical surface mass
density is $\leq 20\%$, for clusters at redshift $z_{c} \sim 0.3$,
when the source redshift varies in the range $0.7 \leq z_{s} \leq 2$.
From the simulated lensed images, B98 obtained the theoretical optical
depth for forming arcs with length-to-width ratios $l/w > 10$.  By
multiplying the lensing optical depth by the density of background
sources with magnitudes $R < 23.5$ mag, as measured by Smail et al.
(1995), they predicted the average number of arcs per cluster with the
above $l/w$ ratio and $R < 21.5$ mag. The predicted numbers of arcs
were compared with the numbers observed in a sample of 16
X-ray-selected clusters from the Einstein Observatory Extended Medium
Sensitivity Survey (EMSS; Le Fevre et al. 1994). B98 found that the
observed number of arcs was higher by an order of magnitude than the
predictions of the (now-standard) $\Lambda$-dominated cold dark matter
($\Lambda$CDM) cosmological model.

This initial study of arc statistics by B98 motivated several
subsequent studies, both theoretical and observational. From the
observational side, improvements in the volume and quality of data
have made the statistics more accurate. Zaritsky and Gonzalez (2003)
measured the frequency of arc occurrence in a sample of clusters with
redshifts $0.5<z_{c}<0.7$, observed as part of the Las Campanas
Distant Cluster Survey (LCDCS). They compared their results for arcs
having $l/w > 10$ and $R < 21.5$ mag with a larger observed sample of
38 clusters from the EMSS (Luppino et al. 1999).  After correcting the
observed incidence of arcs in the EMSS sample for incompleteness, they
found that the arc incidence is lower than, but consistent with, the
arc incidence in the LCDCS sample. Arc statistics were also derived
for the Red-Sequence Cluster Survey (RCS) by Gladders et al.  (2003)
who found the number of arcs to be larger by a factor of $10-20$
than the theoretical value predicted by B98 for clusters in the
redshift range $0<z_{c}<1$. Thus all three of these more recent
observed samples support the conclusion of B98, of an observed
incidence of giant arcs that is large compared to the B98 $\Lambda{\rm
  CDM}$ predictions.

In parallel, several theoretical studies have also readdressed the
problem.  Wambsganss et al. (2004) studied the dependence of the cross
section for arc formation on the lensed source redshift.  They
performed several lensing simulations, each with artificial sources on
a single redshift plane, as in B98.  However, instead of using single
simulated clusters, they lensed their sources through the whole volume
of an N-body simulation. Furthermore, the magnification of the arcs
was used as a proxy for the $l/w$ ratios, instead of a direct
measurement. They found that the cross section is a steep function of
source redshift, and concluded that the order of magnitude problem
reported by B98 can be resolved by adding sources at redshifts other
than $z_{s}=1$ to the simulations. However, Li et al. (2005) have
found that approximating the $l/w$ ratio by the magnification is
incorrect, and that the cross section dependence on source redshift is
shallower than the one found by Wambsganss et al.

Dalal et al. (2004) repeated the B98 ray tracing analysis of
artificial sources using a larger sample of simulated clusters (of
which B98 had used a subset) and compared their results to the
38-cluster EMSS sample (Luppino et al. 1999), but concluded that the
arc statistics data and the $\Lambda$CDM model are consistent. They
explained the order of magnitude discrepancy reported by B98 as a
combination of several smaller effects : 1) The sky density of EMSS
clusters is lower by a factor of 2 than was assumed by B98; 2) the
background source density used by Dalal et al. in their calculations
(based on Fontana et al.  2000) is higher by a factor of 2 than the
one used by B98; 3) the placement of sources at additional redshifts,
other than $z_{s}=1$, yields a higher optical depth than the one
derived in B98.

The artificial clusters used by the above studies represent the
cluster dark matter component only. Adding a mass component,
associated with the dark and baryonic mass of the cluster galaxies, to
the artificial clusters can affect the cross section for forming giant
arcs in several ways (Meneghetti et al.  2000). The critical lines
will curve around the cluster galaxies, thus increasing the cross
section. On the other hand, the galaxies will split some of the long
arcs into shorter arclets, thus decreasing the giant arc cross
section. After studying these two competing effects, Meneghetti et al.
(2000) concluded that the effect of cluster galaxies on the lensing
cross section is negligible, a conclusion also supported by an
analytical study by Flores et al.  (2000). More recently, Meneghetti
et al. (2003) found that the cD galaxy in each cluster does increase
the cross section, but only by up to $\sim 50\%$ for realistic
parameters. Puchwein et al. (2005) have studied the influence of the
intracluster gas on lensing cross section.

Torri et al. (2004) studied the effect of cluster mergers on arc
statistics by following the lensing cross-section of a simulated
cluster with small time steps as the cluster evolves. They concluded
that, during a merger, the strong-lensing cross section can be
enhanced by an order of magnitude in an optimal projection,
potentially providing yet another contribution towards resolving the
problem reported by B98. The effects on arc statistics of dark-energy
with an equation of state different from that of a cosmological
constant have also been studied. (Bartelmann et al. 2003a; Meneghetti
et al. 2004a)

In view of the conflicting observational and theoretical results, in
this paper we revisit arc statistics using several novel approaches. A
discrepancy between models and observations could arise in a number of
ways, such as a larger cluster number density in reality compared to
the results of simulations, a different mass function, or real
clusters that are more efficient lenses than simulated ones. To
isolate the problem (or problems) that have led to the controversy, we
examine here only the arc production efficiency of mass-matched real
and simulated clusters at one particular redshift, $z_{c}=0.2$. Once
this issue is addressed, the other factors, such as cluster number
density, mass function, and redshift evolution, can be studied
separately. To circumvent the question of source properties (density,
redshift, surface brightness), our lensing simulations are based on
real background sources and their photometric redshifts, instead of
artificial sources at a single redshift. For this purpose, we use the
sources in the Hubble Deep Field (HDF; Williams et al. 1996), with
photometric redshifts from the catalog by Fern\'andez-Soto et al.
(1999). To the artificial lensed data, produced by the simulations, we
add observational effects, such as Poisson noise and artificial
cluster galaxies, which we match to the specific observed cluster
sample to which the simulations are compared. We then use an objective
automated algorithm to detect arcs in both the simulated data and in
the real data. Our inclusion of observational effects in the
simulations allows us to compare directly the incidence of arcs in the
simulations and in the observations, without the need to impose
magnitude cuts on the arcs.

Throughout this paper we adopt a $\Lambda {\rm CDM}$ cosmology with
parameters $\Omega_{m}=0.3$, $\Omega_{\Lambda}=0.7$, and
$H_{0}=70~h_{70}~{\rm km~s}^{-1}{\rm Mpc}^{-1}$. The paper's outline is as
follows. The observed sample we analyze is described in $\S2$. $\S3$
provides a detailed description of the lensing simulations. A short
description of the arc-finding algorithm is given in $\S4$. $\S5$
presents the results, and in $\S6$ we summarize the conclusions.

\section{Observed cluster sample}

To obtain a reliable observational estimate of arc statistics, we have
analyzed a well-defined sample of 10 massive clusters observed with
the Hubble Space Telescope (HST) by Smith et al. (2004). The choice of
massive clusters and of deep observations at HST resolution are
intended to maximize the number of detected arcs, and thus reduce the
statistical error.

The sample of Smith et al. (2004) consists of 10 galaxy clusters from
the X-ray Brightest Abell-type Clusters of galaxies (XBACs) catalog
(Ebeling et al. 1996), with $0.171<z_{c}<0.255$, and Galactic
$E(B-V)\leq0.1$. The X-ray $(0.1-2.4$ keV$)$ flux limit of $f_{X}\geq
5.0\times10^{-12}$ erg cm$^{-2}$ s$^{-1}$ applied to this redshift
range implies X-ray luminosities
$L_{X}\geq4.1\times10^{44}h_{70}^{-2}$ erg s$^{-1}$. Assuming the
Reiprich \& B{\" o}hringer (2002) $L_{X}-M_{200}$ relation, where
$M_{200}$ is the mass enclosed within $r_{200}$, (the radius within
which the average density is equal to $200$ times the critical
cosmological density), the observed cluster masses are in the range
$1.7\times 10^{14}\lesssim M_{200}\lesssim 2.6\times 10^{16}~
h_{70}^{-1}M_{\odot}$, with a mean of $M_{200}=1.6\times 10^{15}~
h_{70}^{-1}M_{\odot}$, and a mean $r_{200}$ of $2.2 h_{70}^{-1}~{\rm
  Mpc}$. This range of masses takes into account the uncertainties in
the best fit to the $L_{X}-M_{200}$ relation of Reiprich \& B{\"
  o}hringer (2002), i.e., the range is increased by the sum of of the
uncertainties in $M_{200}$ for the lowest and highest luminosities. The
cluster properties, along with their projected $M(r\leq
390~h_{70}^{-1}{\rm kpc})$, as estimated by Smith et al. (2004) based
on an analysis of weak and strong lensing, are summarized in Table
$1$.

The Smith et al. (2004) sample was observed with WFPC2 on HST using
the F702W filter, with total exposure times of $6500-7800 {\rm s}$.
The individual exposures of each cluster, centered on one of the three
WF CCDs, were combined, dithered, and cosmic-ray rejected using an
IRAF based script (Gal-Yam, Maoz $\&$ Sharon 2002; Sharon 2003). Each
WF CCD is then a $1600\times1600$~pixel image, $1\farcm3$ on a side, with
a pixel scale of $0\farcs05$.  Rows and columns near the image borders,
where spurious features due to the WFPC2 optics may appear, were
excluded from the image.

\begin{table}[!ht]
\begin{center}
\caption{Observed cluster sample}
\smallskip
\begin{tabular}{ccccc}
\tableline
\noalign{\smallskip}
Cluster & $z$ & $L_{{\rm X}}(0.1-2.4{\rm ~keV})$ &
$M(r\leq390 h_{70}^{-1} {\rm kpc})$ & $M_{200}$ \\
 & & [$10^{44}\, h_{70}^{-2}{\rm erg \,s}^{-1}$] & [$10^{14} h_{70}^{-1}M_{\odot}$] &
 $[10^{15}h_{70}^{-1}M_{\odot}]$ \\ 
\noalign{\smallskip}
\hline
\noalign{\smallskip}
Abell 68   & 0.255 & ~5.32 & $3.51$ & $1.20$ \\
Abell 209  & 0.209 & ~8.44 & $1.25$ & $1.60$ \\
Abell 267  & 0.230 & ~8.32 & $2.06$ & $1.59$ \\
Abell 383  & 0.187 & ~4.86 & $2.80$ & $1.12$ \\
Abell 773  & 0.217 & ~7.75 & $4.01$ & $1.52$ \\
Abell 963  & 0.206 & ~6.28 & $2.58$ & $1.32$ \\
Abell 1763 & 0.228 & ~8.88 & $1.66$ & $1.66$ \\
Abell 1835 & 0.253 & 24.68 & $4.62$ & $3.19$ \\
Abell 2218 & 0.171 & ~4.78 & $4.33$ & $1.10$ \\
Abell 2219 & 0.228 & 12.36 & $2.69$ & $2.04$ \\ 
\noalign{\smallskip}
\hline
\smallskip
\end{tabular} 
\end{center}
Note - Luminosities are taken from the XBACs catalog (Ebeling et
al. 1996).
$M(r\leq390 h_{70}^{-1} {\rm kpc})$ is from Smith et al. (2004)
 based on weak and strong lensing. $M_{200}$ is based on the
$L_{X}-M_{200}$ relation of Reiprich \& B{\" o}hringer (2002). All
numbers have been converted to the adopted $\Lambda$CDM cosmology.
\end{table}

\section{Lensing Simulations}

\subsection{Simulated Cluster Sample}

We performed our lensing simulations using the same primary sample of
five artificial clusters used by B98. As described in detail in B98,
these clusters are the most massive clusters found in a cosmological
$\Lambda$CDM ($\Omega_{m}=0.3,\Omega_{\Lambda}=0.7,h_{70}=1$) N-body
simulation, as a part of the ``GIF'' project (Kauffmann et al. 1999).
The GIF simulations adopted the CDM power spectrum by Bond and
Efstathiou (1984) with a shape parameter of $\Gamma = 0.21$, and a
normalization of $\sigma_{8} = 0.90$ (Eke et el. 1996). The
simulations were run with N=256$^{3}$ particles of $2\times10^{10}~
h_{70}^{-1}$M$_{\odot}$ each, in a comoving volume of $201^{3}$
(Mpc$/h_{70})^{3}$.

The clusters were defined by first locating their cores at $z_{c}=0.2$
using a ``friends-of-friends'' group finder.  All the particles within
an Abell radius, $r_{A}=2.14h_{70}^{-1}$ Mpc, around each core were
collected to form a cluster. The masses of the five clusters within
$r_{200}$ (which has a mean of $2$ $h_{70}^{-1}$Mpc) are in the range
$0.8-1.6 \times 10^{15}~h_{70}^{-1}{\rm M}_{\odot}$.  The
three-dimensional mass distribution of each of these clusters was
projected onto each of the three faces of the simulation volume, thus
obtaining three two-dimensional surface mass density fields. The
gravitational deflection angle field of each projection was calculated
on a grid with an angular resolution of $0\farcs88$ (see Bartelmann
$\&$ Weiss 1994, for details). An area of $160''\times160''$,
corresponding to $0.52\times 0.52 \,h_{70}^{-2}{\rm Mpc}^{2}$ in our
adopted cosmology, and surrounding the center of each simulated
cluster, was cut out of the larger grid. In order to match the
resolution of our background source image (the HDF, see $\S3.2$ below)
we linearly interpolated the deflection-angle grid onto a finer grid
with an angular resolution of $0\farcs04$. The five clusters, with
three projections each, comprise our artificial cluster sample. Table
2 summarizes the properties of the simulated clusters and their
projections.  For each projection, we list $\kappa_{max}$, the maximum
of the convergence, defined as $\kappa=\Sigma/\Sigma_{{\rm cr}}$,
where $\Sigma$ is the projected surface mass density, $\Sigma_{{\rm
    cr}}=D_{l}D_{ls}/4\pi GD_{s}$, and $D_{l}$, $D_{s}$, and $D_{ls}$
are the angular diameter distances between the observer and the lens,
the observer and the source, and the lens and the source,
respectively.

The masses of the simulated clusters are well matched to those of the
observed clusters. Although there is some uncertainty in the estimates
of the observed cluster masses, we do not expect the difference
between the mean masses of the two samples to differ by more than a
factor of $\sim 2$.

\begin{table}[!ht]
\begin{center}
\caption{Simulated $z=0.2$ cluster sample}
\smallskip
\begin{tabular}{ccccc}
\tableline
\noalign{\smallskip}
Cluster & $M_{200}$ & \multicolumn{3}{c}{$\kappa_{max}$} \\
 Id & [$10^{15}~h_{70}^{-1}M_{\odot}$] & x-projection & y-projection &
 z-projection  \\
\noalign{\smallskip}
\hline
\noalign{\smallskip}
cj1109 & $1.35$ & 0.615 & 0.650 & 0.562 \\
cj1209 & $1.58$ & 0.890 & 0.942 & 0.677 \\
cj1309 & $1.03$ & 0.814 & 0.680 & 1.224 \\
cj1409 & $0.93$ & 1.234 & 0.605 & 1.168 \\
cj1509 & $0.77$ & 0.330 & 0.806 & 0.393 \\
\noalign{\smallskip}
\hline
\smallskip
\end{tabular}
\end{center}
\centering
Note - $\kappa_{max}$ is the maximum convergence assuming all sources
are at $z_{s}=1$.
\end{table}
  
\subsection{Background source image}

For a background of sources to be lensed by our artificial clusters,
with properties representative of the galaxies lensed by real
clusters, we use the F606W and F814W images of the HDF. The observed
sample, to which we will compare our simulations, was imaged with the
HST WFPC2 F702W filter, centered at 7020 \AA. In order to match the
simulations and the observations, we approximate the F702W filter as
the average of the F606W and F814W filters.  The summed F606W and
F814W exposures of the HDF reach $I_{{\rm AB}}=29.8$ with ${\rm
  S/N}=5$ (Williams et al. 1996), which is deeper by $\sim 4$ mag than
the observed Smith et al. (2004) cluster images to which we compare
our models. Since the lensed arcs typically have magnifications of
order $10$, the HDF suitably represents the sources to be lensed into
such arcs. The $\sim 5\,{\rm arcmin}^{2}$ area of the HDF is too small
to serve as a source background for $z_{c}=0.2$ clusters. We therefore
used segments of the HDF as building blocks with which we created a
larger, $192''\times 192''$, source image with a scale of
$0\farcs04\,{\rm pixel}^{-1}$. Three segments of $1600\times1600$
pixels $(64''\times64'')$ were cut out of the HDF image around the
centers of each of the three WF CCD's.  A mosaic was then built out of
$3\times 3$ blocks, each block randomly selected from the three
building blocks.

Our lensing calculation requires redshift information for each
background source. We therefore created a redshift image, in which the
pixels of each source are assigned the value of the source photometric
redshift.  The redshift image was created as follows. First, we
detected objects in the HDF mosaic using SExtractor (Bertin and
Arnouts 1996), an object detection and photometry program. SExtractor
produces an object catalog and a ``segmentation image'', an image of
the detected objects in which pixels belonging to each object are
assigned the object's catalog number. The values of the pixels
comprising each object in the segmentation image were then replaced
with the object's redshift by cross referencing the detected object
catalog with the HDF photometric redshift catalog of Fern\'andez-Soto
et al. (1999). All but 39 of the 985 objects we detect in the HDF with
$I_{AB}\leq 28$ mag appear in the Fern\'andez-Soto et al. (1999)
catalog. In addition we detect 393 objects with $28<I_{AB}\leq 29.8$
mag, beyond the flux limit of the catalog. The 39 non-catalog objects
with $I_{AB}\leq 28$ were assigned redshifts based on
their magnitudes, by drawing values from the observed distribution of
redshifts for objects in the catalog with the appropriate magnitude.
For the 393 $I_{AB}> 28$ mag objects, the redshift distribution for
the $I_{AB}=28$ bin was used, which is reasonable given the small
change in redshift distribution beyond $I_{AB}\approx 26$
(Fern\'andez-Soto et al. 1999).

\subsection{Ray tracing method}   

In the small-angle, weak-field, and thin lens regime (e.g., Bartelmann
2003b) lensing is described by the lens equation,
\begin{equation}
\vec{\beta}=\vec{\theta}-\frac{D_{ls}}{D_{s}}\vec{\alpha}(\vec{\theta})\,,
\end{equation}
where $\vec{\beta}$ is a two-dimensional vector describing the angular
position of the source on the sky, $\vec{\theta}$ is the angular
position of a lensed image, and $\vec{\alpha}(\vec{\theta})$ is
the deflection angle due to the gravitational potential of the
cluster.

In previous arc statistics studies, artificial sources on a source
plane at a single redshift were lensed through a lens plane
(e.g., Bartelmann et al. 1998; Meneghetti et al. 2003; Dalal et al.
2004).
Starting from the observer, rays were shot back to every position
$\vec{\theta}$ in the lens plane. Each ray was then deflected towards
a position $\vec{\beta}$ according to the lens equation.  Only if an
artificial source existed at $\vec{\beta}$, would the source be mapped
onto the lens plane. Thus, a ``lensed image'' of the source plane in
the lens plane was produced.

In the present work, as opposed to using artificial sources at a
single redshift, we use real background sources, each at its
photometric redshift. Since $D_{ls}/D_{s}$ is no longer constant for a
given lens redshift, a single value for the position $\vec{\beta}$
cannot be obtained, given particular values of $\vec{\theta}$ and
$\vec{\alpha}(\vec{\theta})$. Instead, the lens equation, which
depends linearly on $D_{ls}/D_{s}$, produces a range of values for
$\vec{\beta}$. To account for this, we calculate the maximum and
minimum values of $D_{ls}/D_{s}$ for the redshift range spanned by the
sources in the redshift catalog (note that $D_{ls}/D_{s}$ is not
necessarily a monotonic function of $z_{s}$). By introducing these
values into the lens equation, we obtain $\vec{\beta}_{1}$ and
$\vec{\beta}_{2}$, respectively. The vector
$\vec{\beta}_{2}-\vec{\beta}_{1}$ define a line on the source plane
tracing the different positions $\vec{\beta}(z_{s})$ corresponding to
different source redshifts. We then scan the line for pixels belonging
to detected sources. When such a pixel, belonging to a certain source
at $z_{s}$, is found, we calculate the $\vec{\beta}$ corresponding to
$z_{s}$. If the observed and the calculated $\vec{\beta}$'s match, we
lens the source pixel by adding its count rate to the pixel at
position $\vec{\theta}$ in the lens image. The count rate in every
pixel in the simulated lensed HDF image is multiplied by the mean
exposure time of each of the observed clusters, $7430$~s.

Ten lensing realizations were performed for every cluster projection,
where in each realization the source image was randomly shifted in
relation to the center of the lens.  The five simulated clusters,
times three projections, times 10 positions relative to the
background, thus produce a sample of 150 simulated F702W lensed
images.

\subsection{Observational effects}

The next step in our study, which again extends upon previous work, is
to apply to the simulations all of the observational effects that
exist in the real data to which we will compare these simulations (see
Meneghetti et al. 2004b, for a first attempt at applying observational
effects to such simulations). These observational effects include
light from the foreground cluster galaxies, sky and detector
backgrounds, Poisson noise, and readout noise.

The light of simulated cluster galaxies was added based on the
luminosity function of early type galaxies in clusters derived by Goto
et al.  (2002). They identified clusters with $0.02< z_{c}<0.25$ based
on colors and galaxy over-density in the Sloan Digital Sky Survey
(SDSS), and calculated the Schechter (1976) luminosity function
parameters $M^{\ast}$ and $\alpha$ in the SDSS $u,~g,~r,~i,~{\rm
  and~}z$ bands for early- and late-type galaxies.  Goto et al. (2002)
used three different criteria for distinguishing between the
early-type and the late-type galaxies, resulting in three different
luminosity functions.  We used the parameters $M_{{\rm
    AB}}^{\ast}=-22.1$ mag and $\alpha=-0.75$ which are the averages
of the parameters obtained in the $r$, and $i$ bands for early-type
galaxies identified as such based on their light concentration.

We determined the number of galaxies to be added to our simulated
cluster images using the properties of our observed cluster sample.
We first normalized the luminosity function so as to have $21.5$
galaxies in the absolute AB magnitude range $-22.7\leq M_{{\rm
    F702}}\leq-20.7$, matching the average number of galaxies per
cluster in this absolute magnitude range that we find in the observed
sample. The normalized luminosity function then has 67 galaxies in the
range $-23.2\leq M_{{\rm F702}}\leq-16.7$. We then added to each
simulated cluster image 67 galaxies, with absolute magnitudes drawn
from the luminosity function.

The artificial cluster galaxies were placed in each simulated cluster
image at a position drawn at random from a distribution following the
surface mass density of the cluster. In addition, we placed an
artificial cD galaxy at the position of the surface mass density peak,
with a random AB magnitude in the range $-24\leq M_{{\rm
    F702}}\leq-23.2$, also based on the range seen in the observed
sample. The light of each galaxy was distributed with a S\'ersic
projected radial profile,
\begin{equation}
\log\left(\frac{I}{I_{0}}\right)=-b_{n}\left[(R/R_{e})^{\frac{1}{n}}-1\right]\,,
\end{equation}
where $I$ is the surface brightness, $R_{e}$ is the half-light radius,
and $I_{0}=I(R_{e})$, with a random axis ratio in the range $[0.5,1]$,
and with a random orientation. $R_{e}$ for each galaxy was calculated
using the relation between luminosity and half-light radius
$L_{i}\propto R_{e}^{1.59}$ obtained from Fig. 5 of Bernardi et al.
(2003), while the other S\'ersic profile parameters were taken to be
\begin{eqnarray}
\log n&=&0.28+0.52\log R_{e}\nonumber\,,\\
b_{n}&=&0.868n-0.142\,,
\end{eqnarray}
following Caon, Capaccioli, and D'Onofrio (1993), with $R_{e}$ in kpc.
The surface brightnesses of these simulated $z=0.2$ galaxies were
converted to counts pixel$^{-1}$, using the WFPC2 F702W sensitivity
(Baggett et al. 2002) and the mean exposure time of the observed
sample.

A uniform background was added to all the simulated lensed images,
based on the average background value of $186\, e^{-}$ pixel$^{-1}$ in
the observed cluster sample, scaled to the pixel size of the simulated
sample. Poisson noise was then added to each pixel in the image
according to the total counts (contributed by background, cluster
galaxies, and lensed background objects).  Finally, readout noise of
$1.4\, e^{-}$ was added to each pixel, corresponding to the readout
noise per original pixel of $5.2\,e^{-}$, times $\sqrt{3}$ (each
cluster exposure was split into 3 sub-exposures) and divided by the
6.25 dithered sub-pixels comprising a real pixel.

Two of the observed clusters (Abell 2218, and Abell 2219) have a
background level which is lower by $\sim 40\%$ than that of the other
clusters. In order to study the effect of adding low background to our
images, we produced a secondary sample of $90$ lensed data sets with a
low background level of $71\,e^{-}$ per pixel. The implications of
using a lower background are discussed in $\S 5.1$.

Fig. 1 shows several examples of the simulations, before
and after the addition of observational effects.
\begin{figure}[!ht]
\centering
\subfigure[cluster cj1209]{
\includegraphics[width=8cm , angle=0]{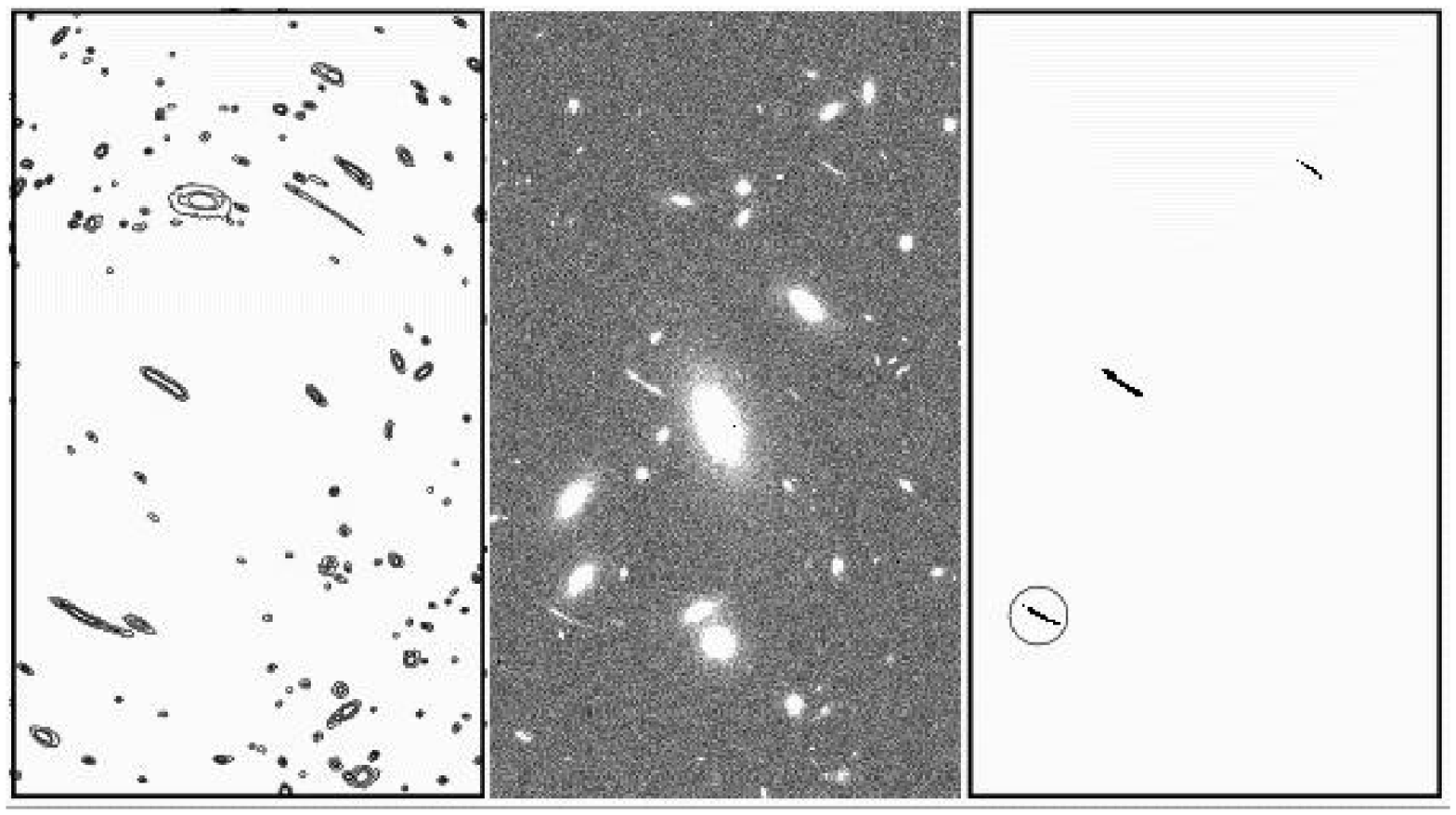}}
\subfigure[cluster cj1309]{
\includegraphics[width=8cm , angle=0]{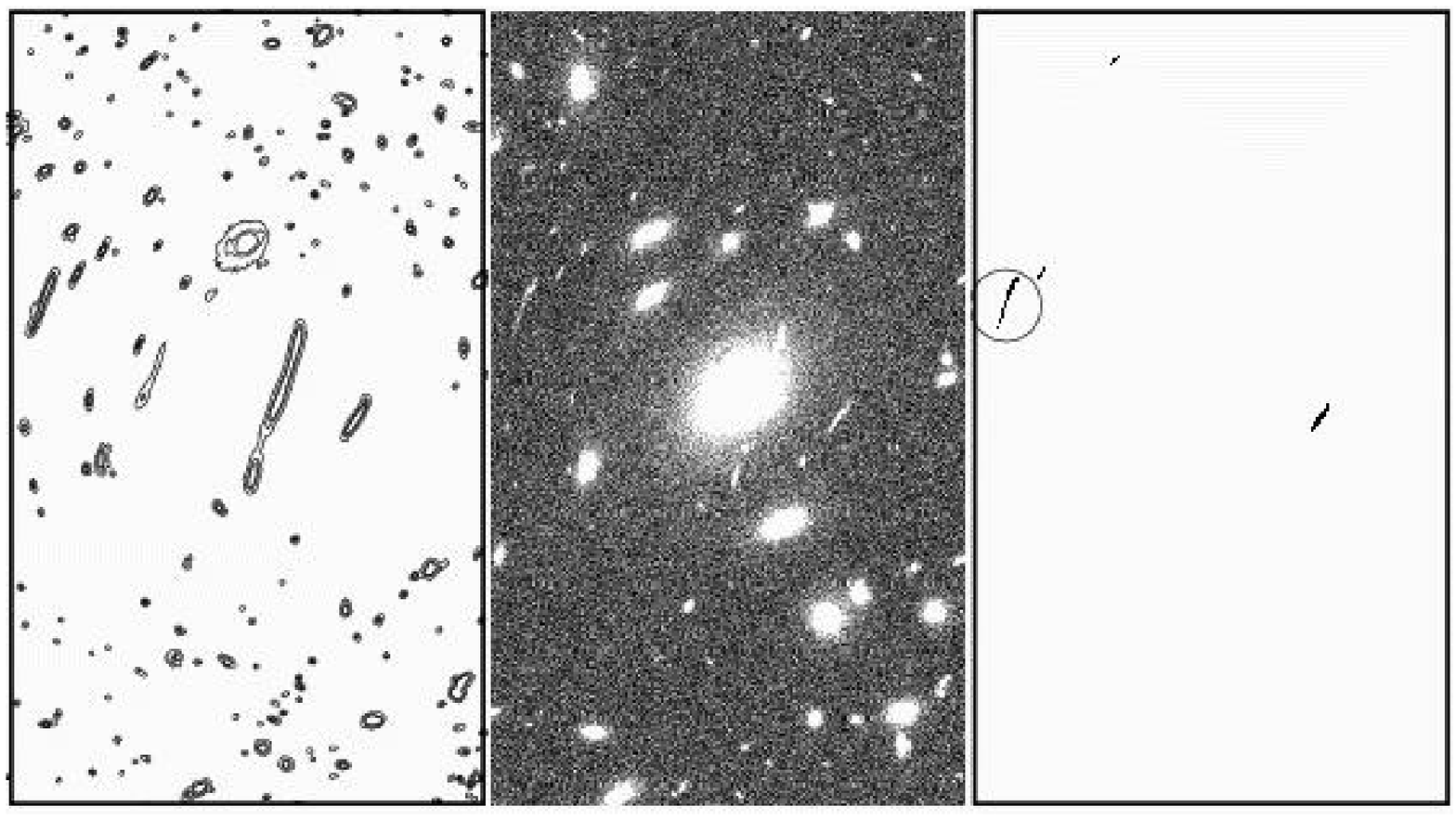}}
\subfigure[cluster cj1409]{
\includegraphics[width=8cm , angle=0]{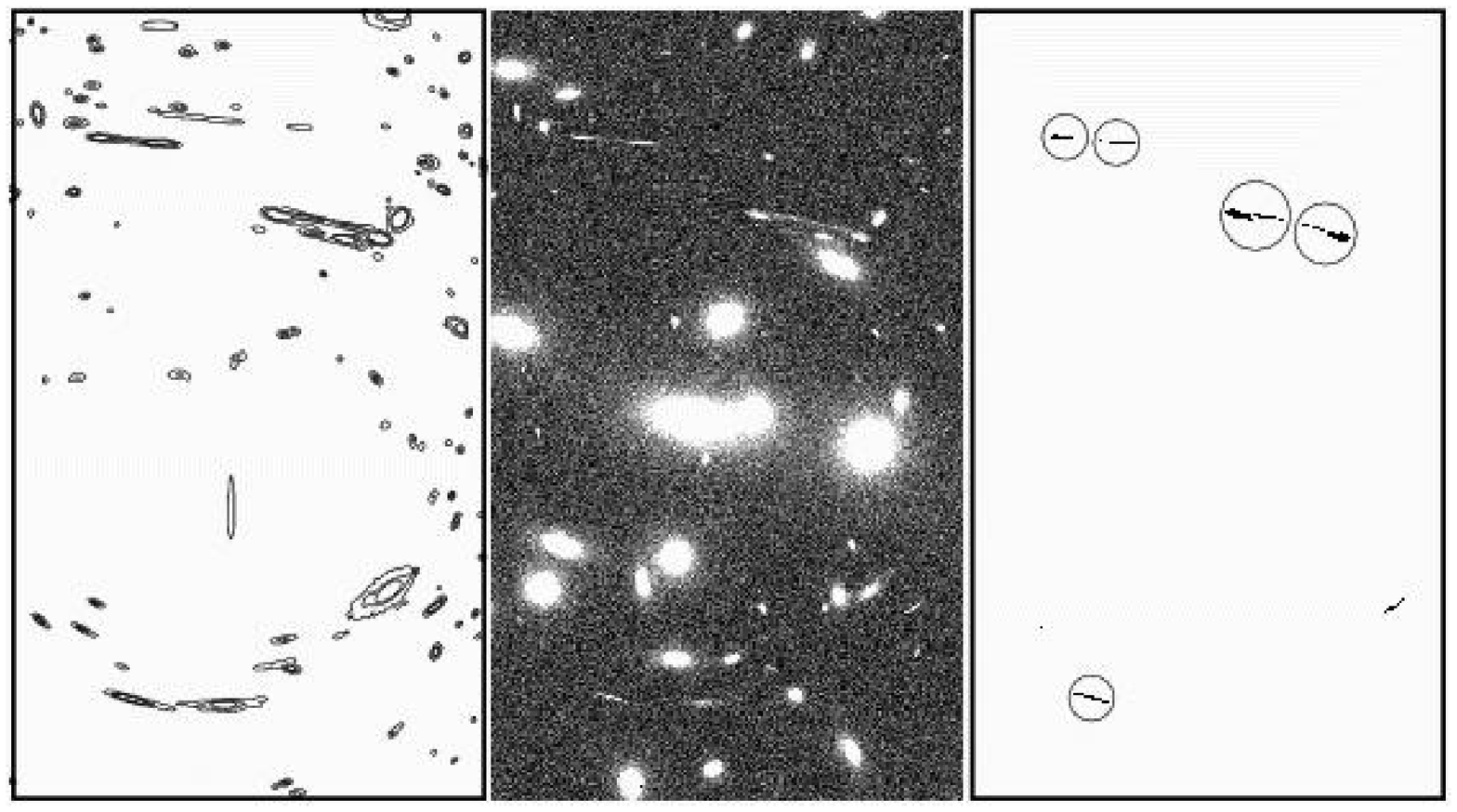}}
\caption{Examples of simulated lensed images. The left panels show a contour map
  of the lensed image produced by the lensing simulations. The middle
  panels show the lensed images after the observational effects are
  added. The right panels show the arcs detected by our arc-finding
  algorithm with $l/w \geq 7$. Giant arcs $(l/w \geq 10)$ are
  circled.}
\end{figure} 

\section{Arc Detection Algorithm}

In order to objectively compare the arc statistics of simulated and
observed clusters, we have devised an automatic arc finding program
that can be applied to both real and simulated data. Our
algorithm\footnote{available at http://wise-obs.tau.ac.il/$\sim$assafh} is
implemented in a simple script that makes repeated calls to the
SExtractor (Bertin \& Arnouts 1996) object finding program (full
details are given in Appendix A). Briefly, in each SExtractor call,
objects are detected using slightly different detection parameters.
Objects with an axis ratio below some threshold are then removed from
the segmentation images produced in the different calls. Finally, arcs
are detected in an image produced by combining the above segmentation
images. The arc length-to-width ratio, following Miralda-Escude
(1993), is measured as follows. First, we find the center of the arc,
defined as the pixel with the maximum flux. Then, the point in the arc
farthest from the center is located, and finally the farthest point
from the second point is located. The arc length is defined as the sum
of the distances from the arc center to the two extreme points, and
the width of the arc is defined as its area divided by its length. The
performance of the algorithm on several of the simulated clusters is
illustrated in Fig. 1. In Appendix A, we also compare our arc-finding
algorithm to a similar tool presented by Lenzen et al. (2004).


\section{Arc Statistics}

\subsection{Results for the simulated sample}

Applying our arc finder to the 150 simulated lensed images, we detect
a total of $141$ giant $(l/w\geq 10)$ arcs. The average number of arcs
per cluster is $\bar{N}=0.94^{+0.08}_{-0.08}$. Almost all of the arcs,
$136$ out of $141$, were produced by three out of the five simulated
clusters. Fig.  2 shows the number of arcs detected in the different
background source realizations for each of the cluster projections.
Apart of the inefficiency of arc production by two of the clusters, we
see that the efficiency of the other clusters sometimes depends
strongly on projection axis (see also Dalal et al. 2004). Not
surprisingly, comparison of Figure $2$ and Table $2$, shows that the
most efficient cluster projections tend to have high values of
$\kappa_{max}$, and vice-versa.

In the previous studies by B98 and Dalal et al. (2004), observational
effects were not included in the simulations, and therefore
observations and simulations were compared only for arcs that are
luminous in addition to being giant. Wu \& Hammer (1993) defined giant
luminous arcs as having $l/w > 10$ and $B<22.5$ mag (or $R<21.5$ mag; B98).
If we adopt an analogous definition of $m_{{\rm F702}}\leq 21.8$ mag
for luminous arcs, the total number of giant luminous arcs we detect
is $33$, or $\bar{N}=0.22^{+0.05}_{-0.04}$ arcs per cluster.
\begin{figure}[!ht]
\centering
\includegraphics[width=6cm , angle=-90]{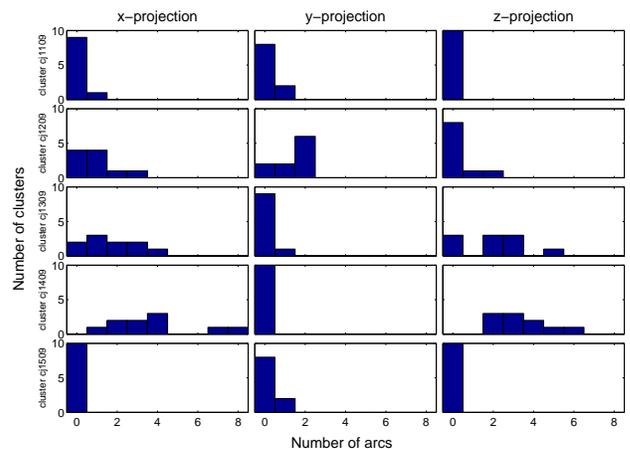}
\caption{Distribution of the number of detected arcs per cluster with
  $l/w > 10$ in the simulated sample.  Each row represent a different
  cluster and each column represent a different projection. The first
  and last clusters are inefficient arc producers, and one of the
  projections of the third and fourth clusters is also inefficient.}
\end{figure}

To investigate the effect of using a realistic source redshift
distribution, we have repeated our simulations with all sources at
redshift $z_{s}=1$, as in B98. The total number of arcs and giant
luminous arcs in our standard simulations, $141$ and $33$, changes by
only $-16$ and $+11$, respectively, consistent to within $2\sigma$
with Poisson fluctuations. The consistency is also evident when
comparing the number distribution of arcs per cluster in the two sets
of simulations, as shown in Fig. 3. In the Fern\'andez-Soto et al.
(1999) catalog, the number of bright objects per arc-minute peaks near
$z=1$, and the median redshift of bright objects is also $1.0$. In
order to investigate the possibility that the HDF has an atypical
redshift distribution due to cosmic variance, we have repeated the
simulations, but with all sources at $z_{s}=1.5$. We detect a total of
$168$ arcs and $50$ giant luminous arcs, which is an increase of
$19\%$ and $51\%$, respectively, compared to the number of arcs we
detect in our standard simulations. These results are also shown in
Fig 3. Thus, there will be a small or modest increase in the predicted
number of arcs if in the HDF the characteristic redshifts are
anomalously low. Similarly, an atypically low space density of sources
in the HDF (e.g. Dickinson 2000), would underpredict the number of
arcs.
\begin{figure}[!ht]
\centering
\includegraphics[width=6cm, angle=-90]{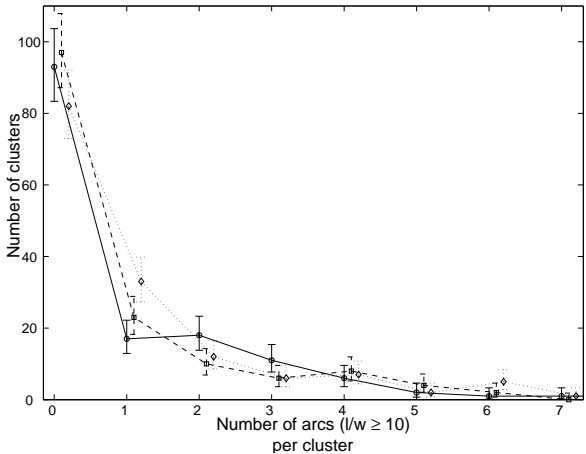}
\caption{Distribution of detected arcs per cluster with
  $l/w\geq 10$ in the simulations, for different assumed source
  redshifts. Circles and solid lines represent the simulations in
  which sources with real photometric redshifts were lensed. Squares
  and dashed lines result when all sources are assigned a redshift
  $z_{s}=1$. Diamonds and dotted lines are for all sources at
  $z_{s}=1.5$. Poisson error bars are shown.  Here and in subsequent
  figures, a slight offset in the horizontal direction has been
  induced between the curves, for the sake of clarity}
\end{figure}


The observational effects included in our simulations influence the
detection process in several ways. Image depth and the local
background determine whether a lensed source will be detected.
Furthermore, since the lensed sources have real light profiles rather
than a constant surface brightness, it is possible that only certain
parts of the lensed source are detected. As a result, the $l/w$ ratio
of an arc with a real light profile, as detected in an image to which
observational effects have been added, will sometimes differ from the
$l/w$ of the same arc in a noise-free image produced by lensing a
source with a constant surface brightness profile. In addition, the
cluster galaxies also affect the results of the detection process. An
arc, or part of it, which lies close to a cluster galaxy will
sometimes be misclassified as part of that galaxy. The arc can also be
broken into two separate segments if a galaxy is positioned on top of
it.

We also studied the implications of introducing a lower background
into our simulated images. Applying our arc-finding program to the
secondary sample of images, mentioned in $\S3.4$, we find that the
detection process is affected in several ways: 1) In the lower
background images, wider parts of the arcs are detected, thus changing
their $l/w$, sometimes below or above the detection threshold of $l/w
\geq 10$.\\ 2) An arc which is detected as two separate arcs in the
high-background image can be detected as one continuous arc in the
low-background image (see, e.g., Fig. 4). These competing effects seem
to largely cancel, as the number of giant arcs and giant luminous arcs
in the low-background images was higher by $11$ and lower by $3$,
respectively, than their number in the higher-background images. A
systematic study of the influence of these and other observational
effects will be presented in Meneghetti et al.  (2005, in
preparation).
\begin{figure}[!ht]
\centering
\plotone{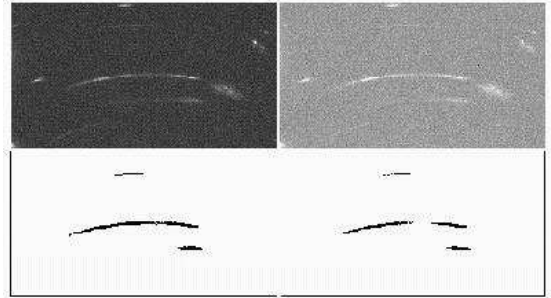}
\caption{Examples of simulated lensed images with low background (top left) and high
  background (top right). The bottom panels show the arcs detected in
  each case. A higher background can lead to the apparent break-up of
  long arcs. On the other hand, a low background can reveal low
  surface-brightness regions of an arc, leading to an assignment of a
  lower $l/w$ ratio.}
\end{figure}

\subsection{Results for the observed sample and comparison with the simulations}

We applied the same arc-finding algorithm described in $\S4$ to the
observed sample. Fig. 5 shows the relevant image sections and the arcs
identified by the arc finder. A total of 12 arcs with $l/w > 10$ are
detected in 7 out of the 10 observed clusters, giving a mean of
$1.20^{+0.46}_{-0.34}$ arcs per cluster. Only two of these arcs are
giant luminous arcs, i.e., a mean of $0.20^{+0.26}_{-0.13}$ giant
luminous arcs per cluster. Before comparing these results to the
results of the simulated sample we take into consideration the fact
that the area in which the observed arcs were detected, covered by the
WFPC2 CCDs, is only a fraction of the total cluster area.  Based on
the shape of the WFPC2 mosaic, we have calculated the average
effective coverage area as a function of the distance from the cluster
center. We then assign a detection probability to each detected arc in
the simulated sample based on the above function, obtaining
$0.71^{+0.06}_{-0.06}$ arcs per cluster and $0.19^{+0.04}_{-0.03}$
giant luminous arcs per cluster. Our simulated results, both for the
total number of giant arcs and for the number of giant luminous arcs,
are therefore consistent with the observations to within $1.6\sigma$.
Examining the observed and the simulated distributions of arc number
per cluster, a Kolmogorov-Smirnov test indicates that the hypothesis
that both distributions are drawn from the same parent distribution
can be only marginally rejected, with a probability of $96.7\%$. The
difference is in the sense that real clusters produce a flatter
distribution -- fewer clusters are arcless and more clusters have two
giant arcs. As seen in Fig. 6, the small number of observed clusters
dominates the uncertainty in the comparison. Larger observed samples
can therefore allow more stringent tests.
\begin{figure*}[!ht]
\centering

\subfigure[Abell 68]{
\includegraphics[width=7cm, angle=0]{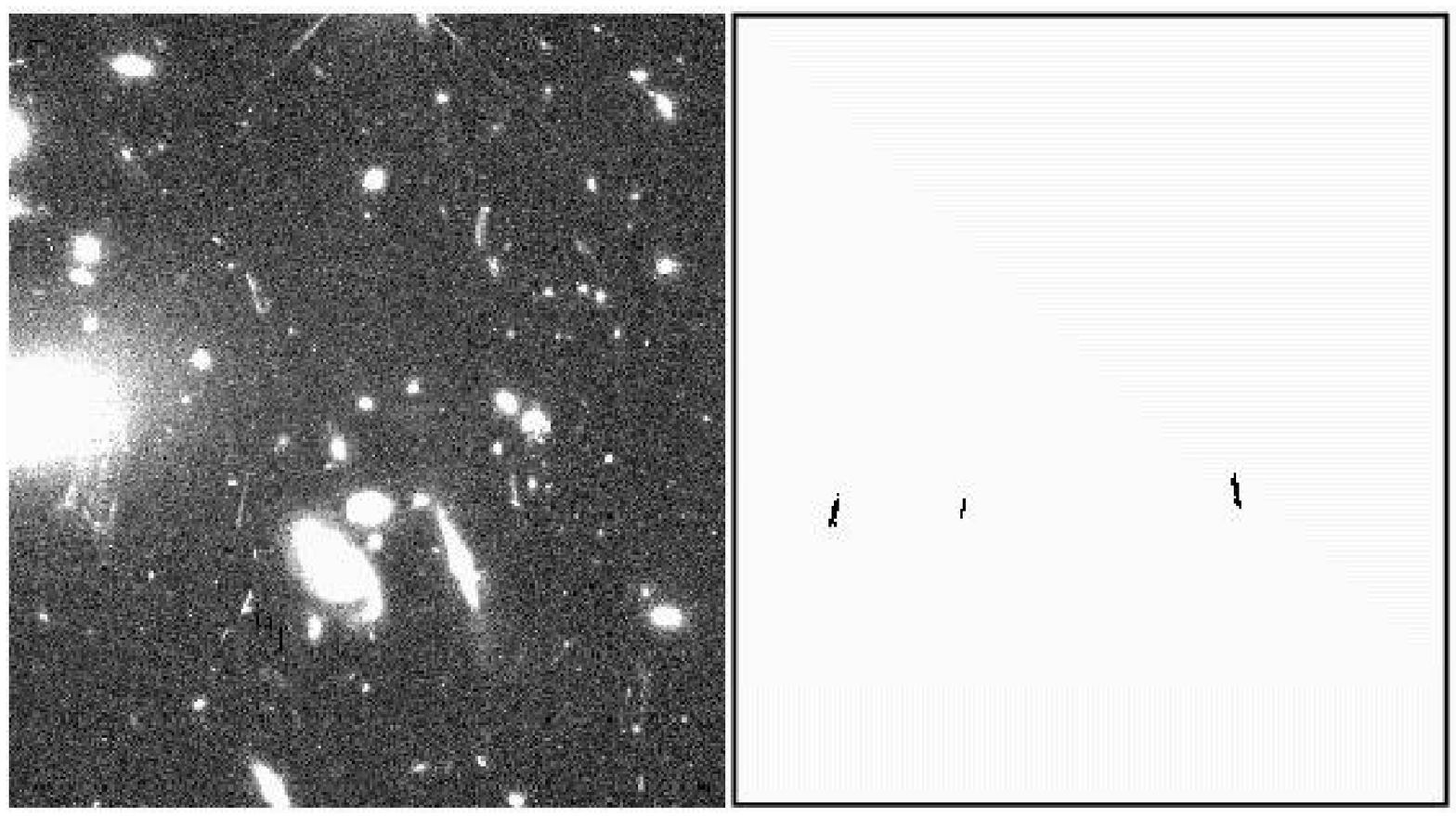}}
\hspace{0.5cm}
\subfigure[Abell 68]{
\includegraphics[width=7cm, angle=0]{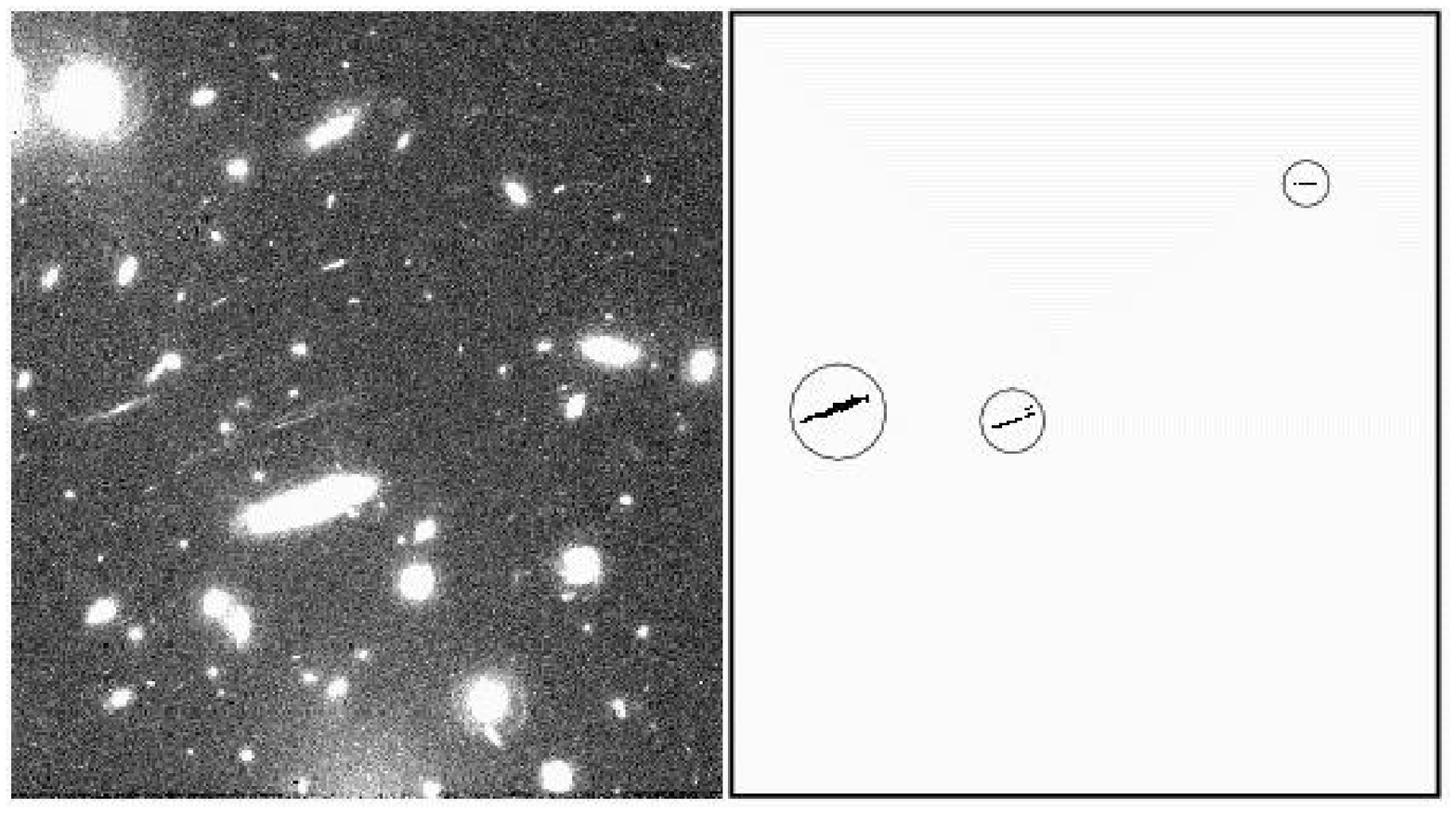}}
\subfigure[Abell 209]{
\includegraphics[width=7cm, angle=0]{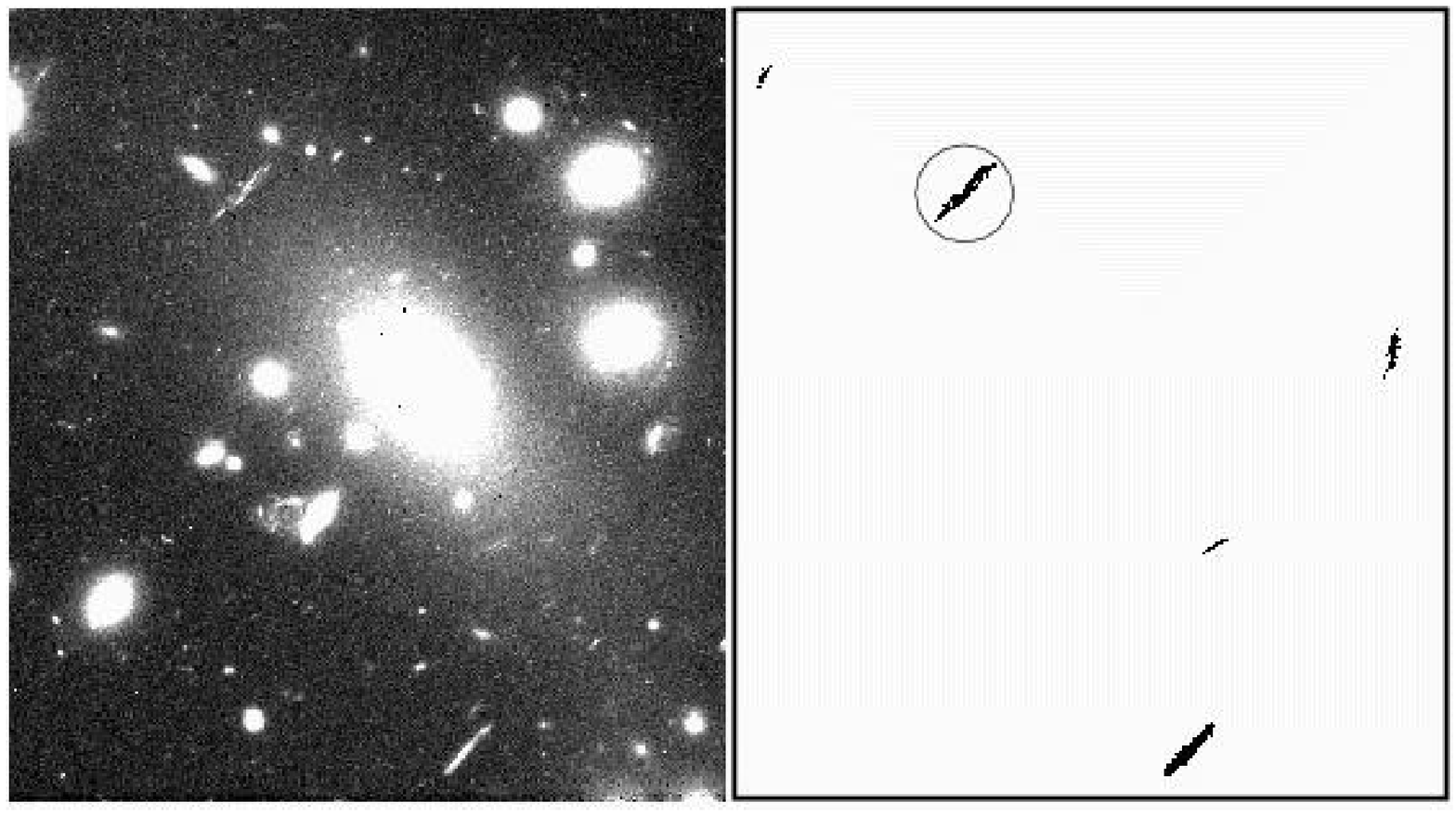}}
\hspace{0.5cm}
\subfigure[Abell 267]{
\includegraphics[width=7cm, angle=0]{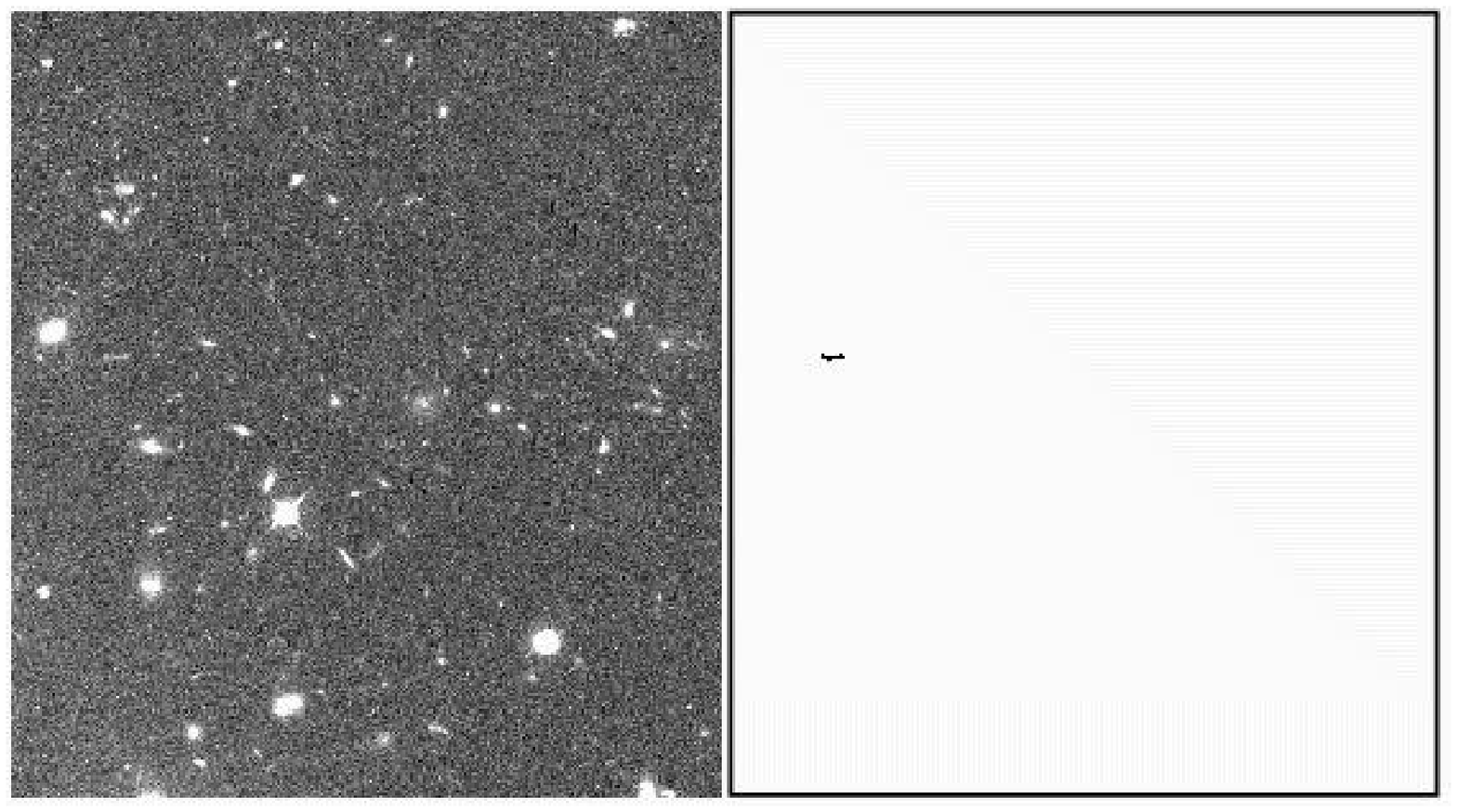}}
\subfigure[Abell 383]{
\includegraphics[width=7cm, angle=0]{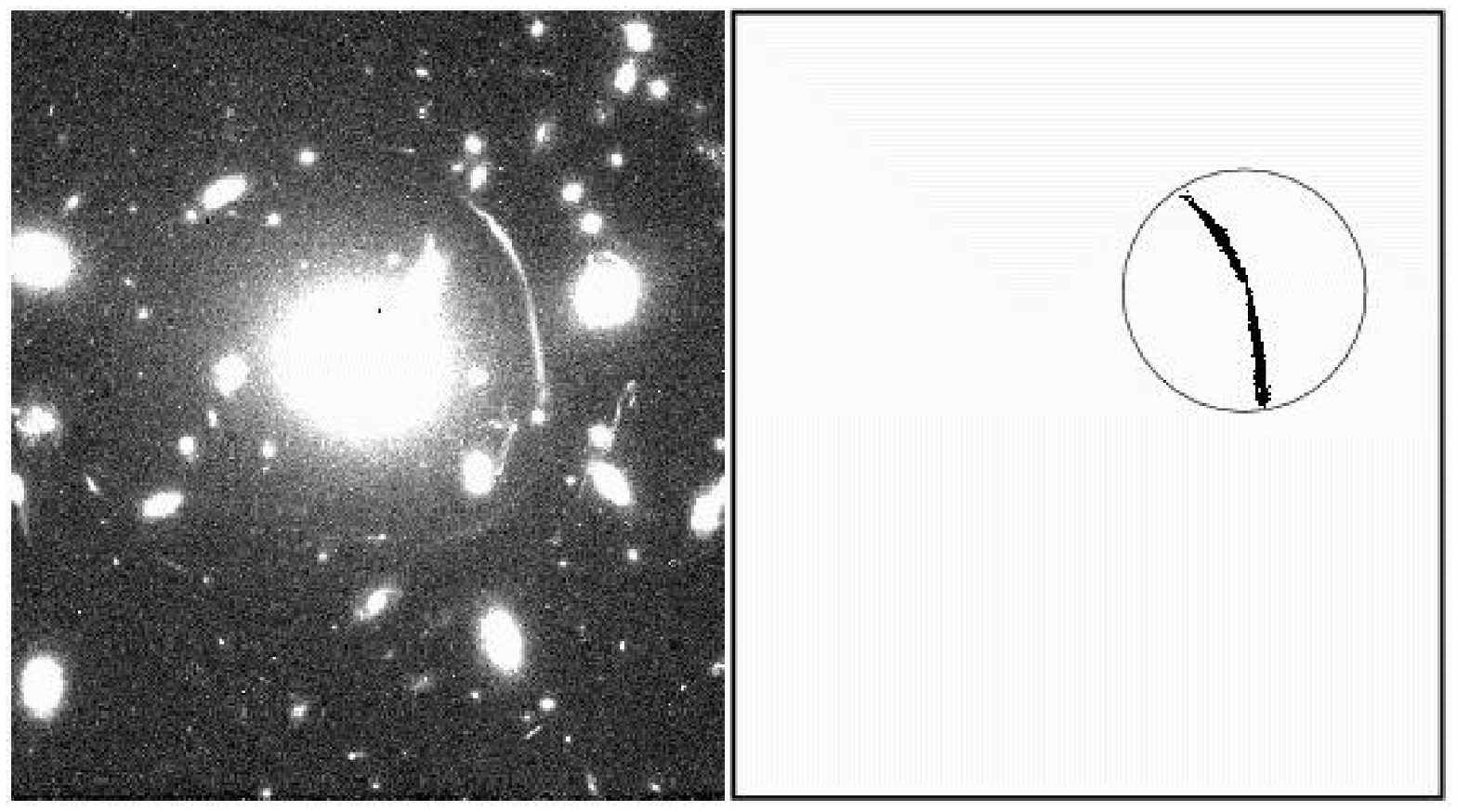}}
\hspace{0.5cm}
\subfigure[Abell 1763]{
\includegraphics[width=7cm, angle=0]{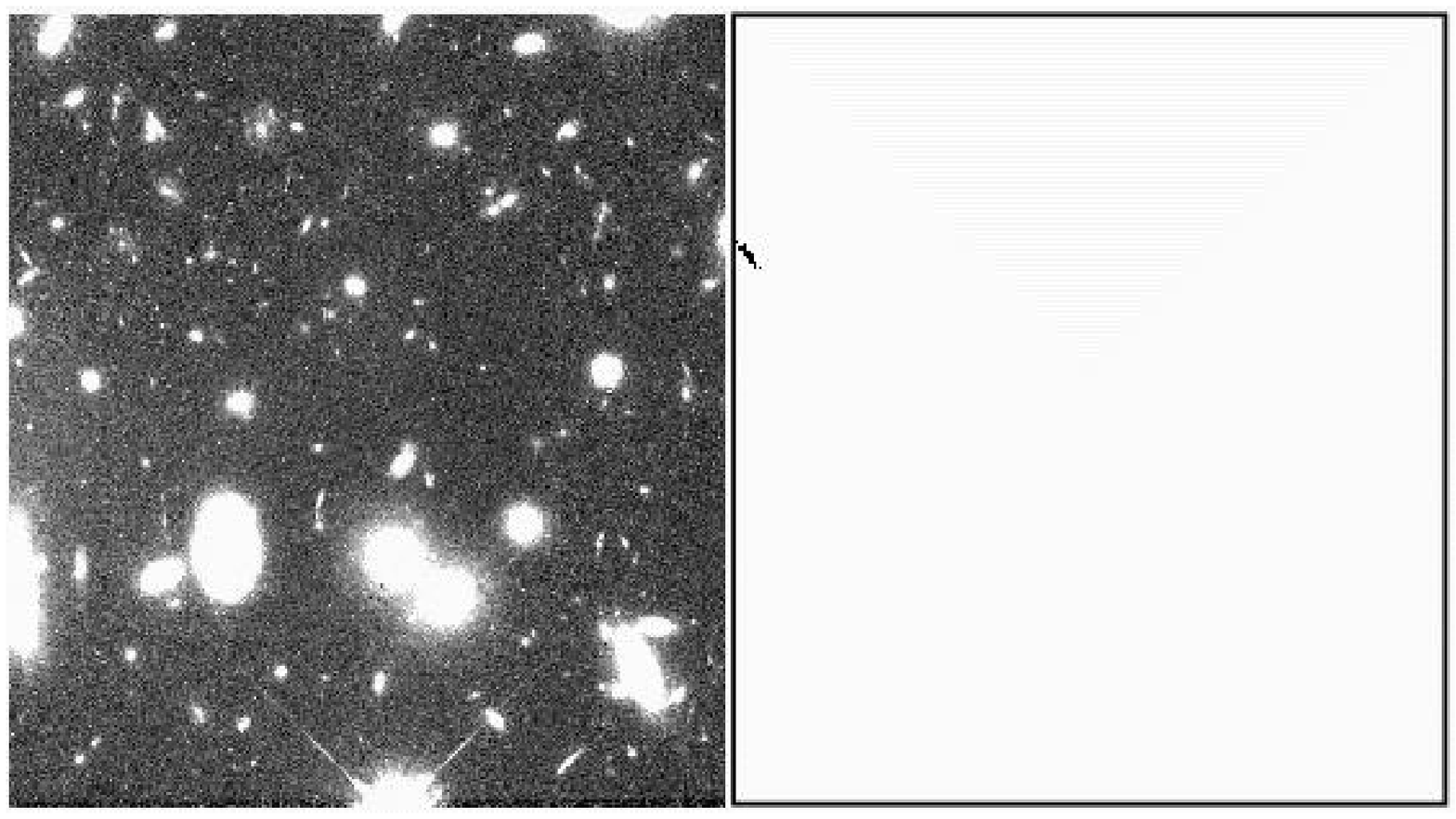}}
\subfigure[Abell 1835]{
\includegraphics[width=7cm, angle=0]{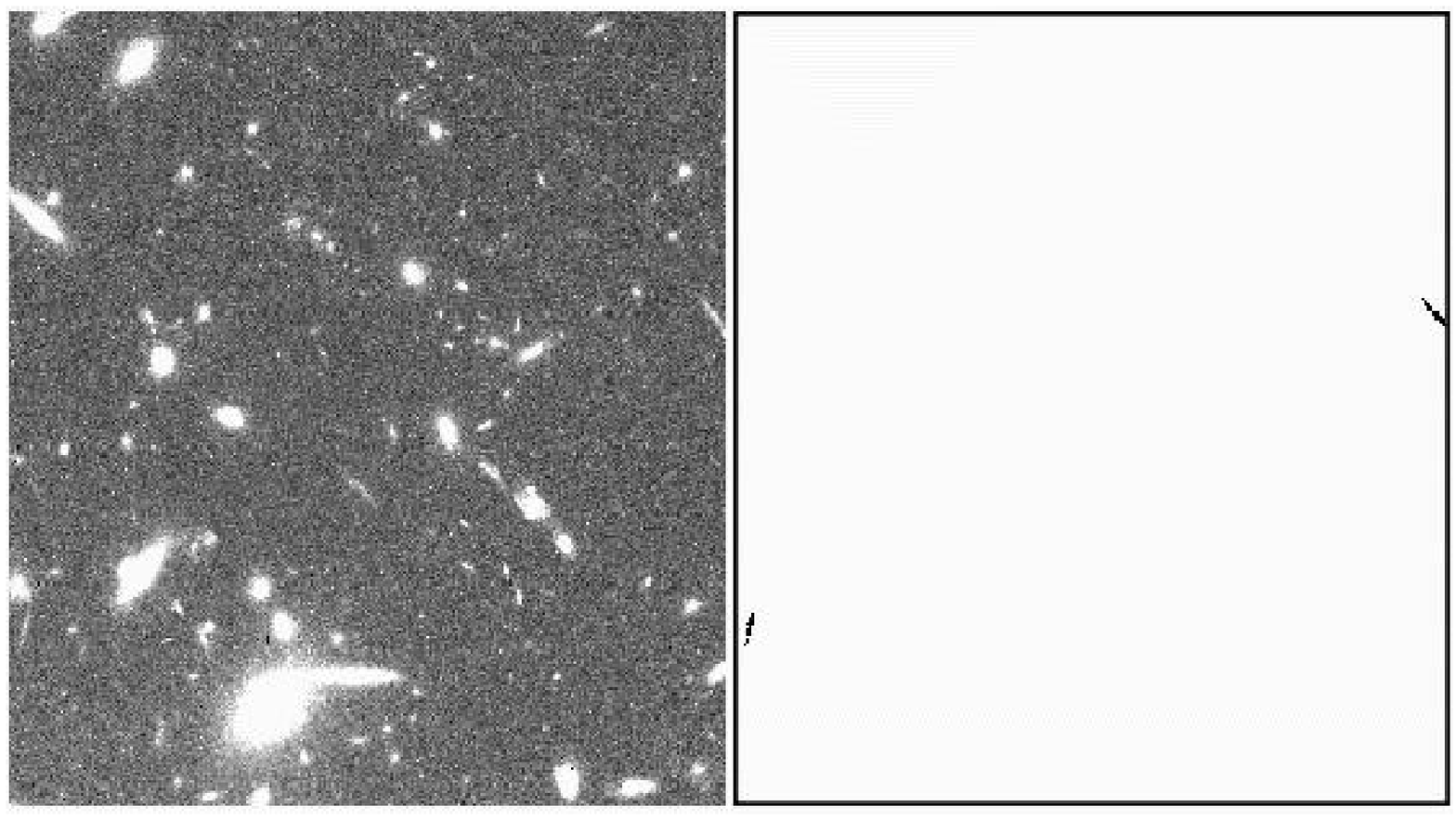}}
\hspace{0.5cm}
\subfigure[Abell 2219]{
\includegraphics[width=7cm, angle=0]{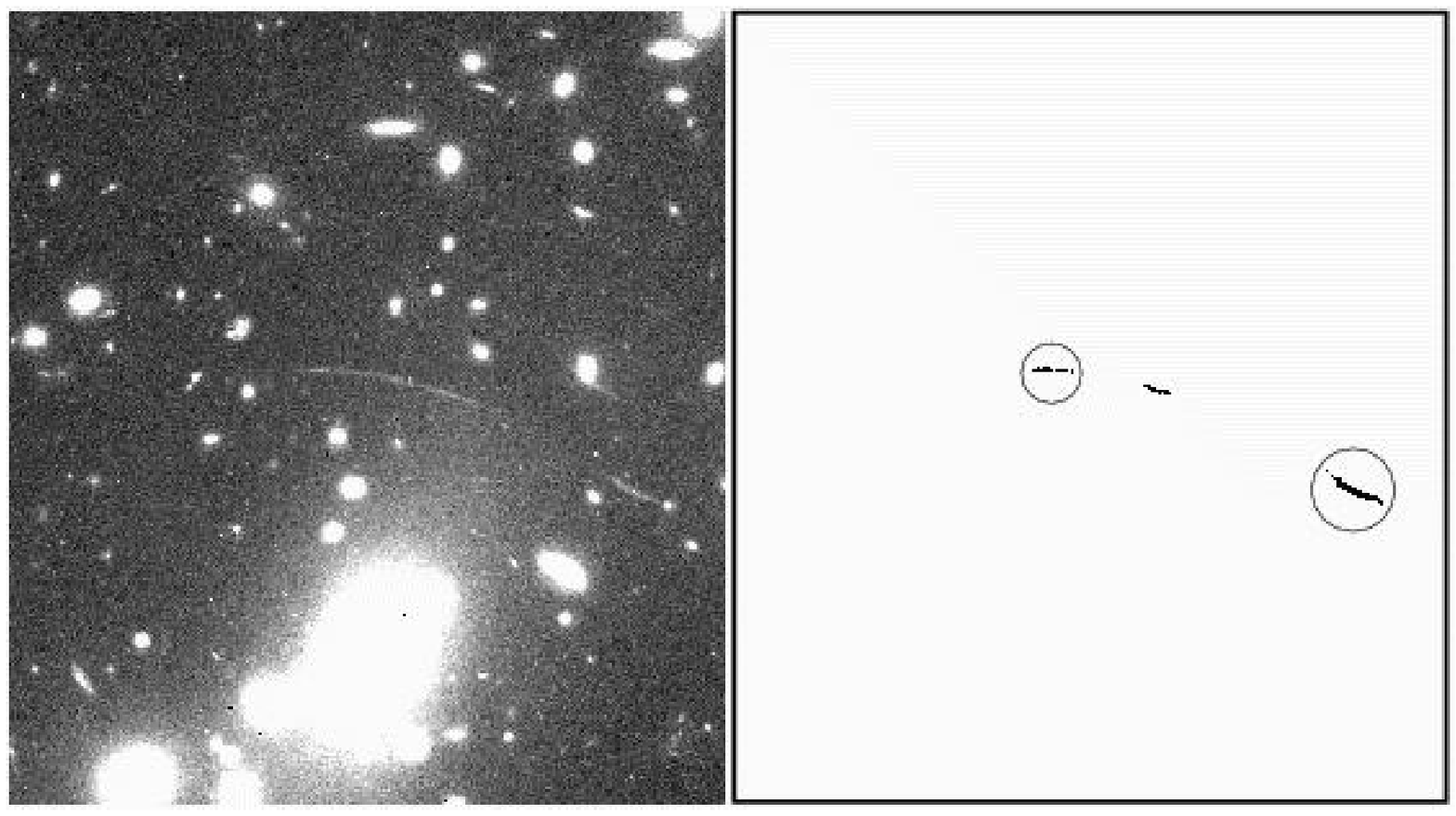}}
\vspace{0.5cm}
\caption{$62'' \times 69''$ sections of WFPC2 images of the observed
  clusters with arcs, showing only
  the regions in which arcs are detected. The right hand panels show
  the arcs (with $l/w \geq 7$) detected in each case by our automated arc-finding
  algorithm. Giant arcs ($l/w \geq 10$) are circled.}
\end{figure*}

\addtocounter{figure}{-1}

\begin{figure*}[!ht]
\centering
\addtocounter{subfigure}{+8}
\subfigure[Abell 773]{
\includegraphics[width=6.5cm, angle=0]{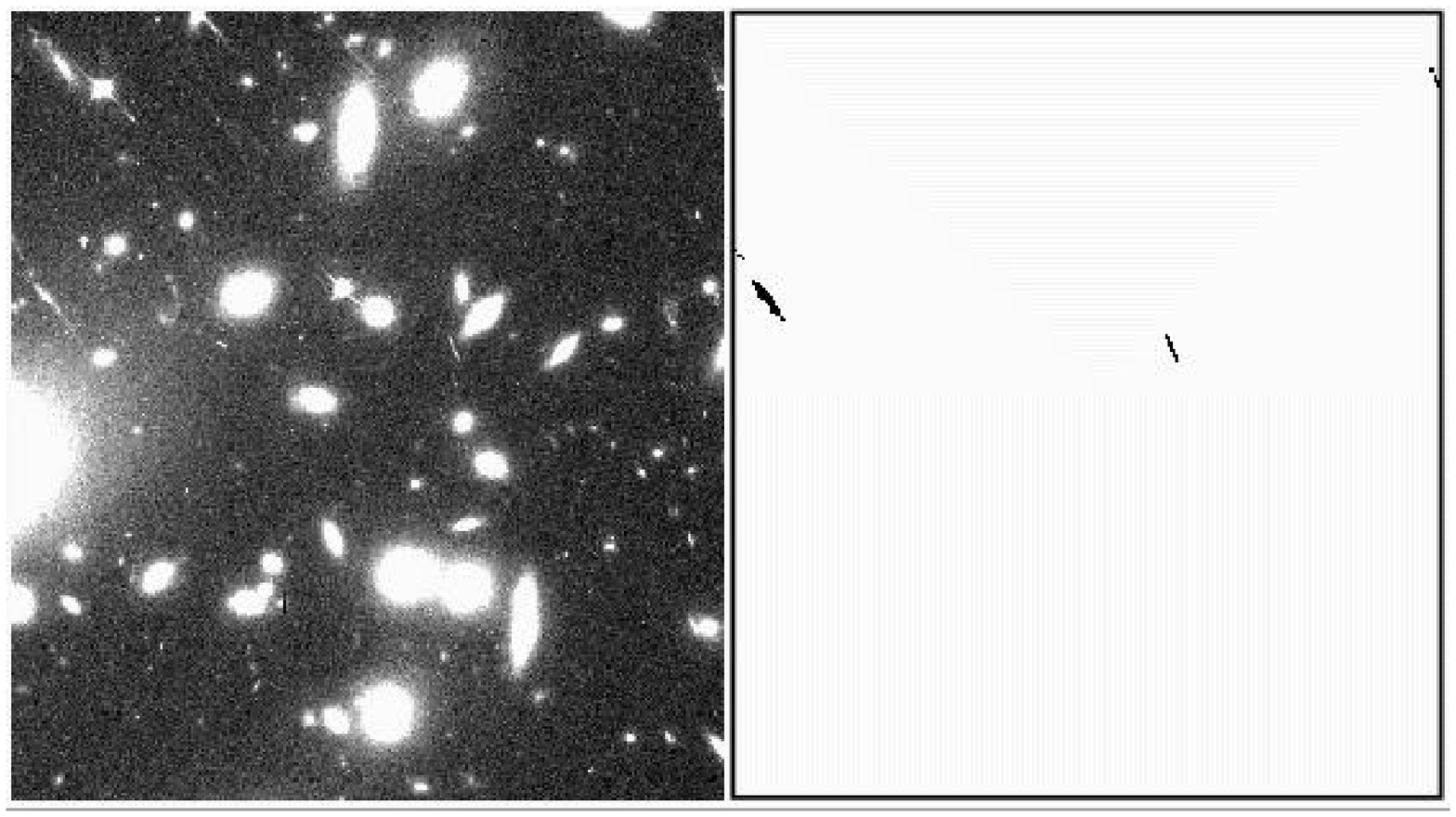}}
\hspace{0.5cm}
\subfigure[Abell 773]{
\includegraphics[width=6.5cm, angle=0]{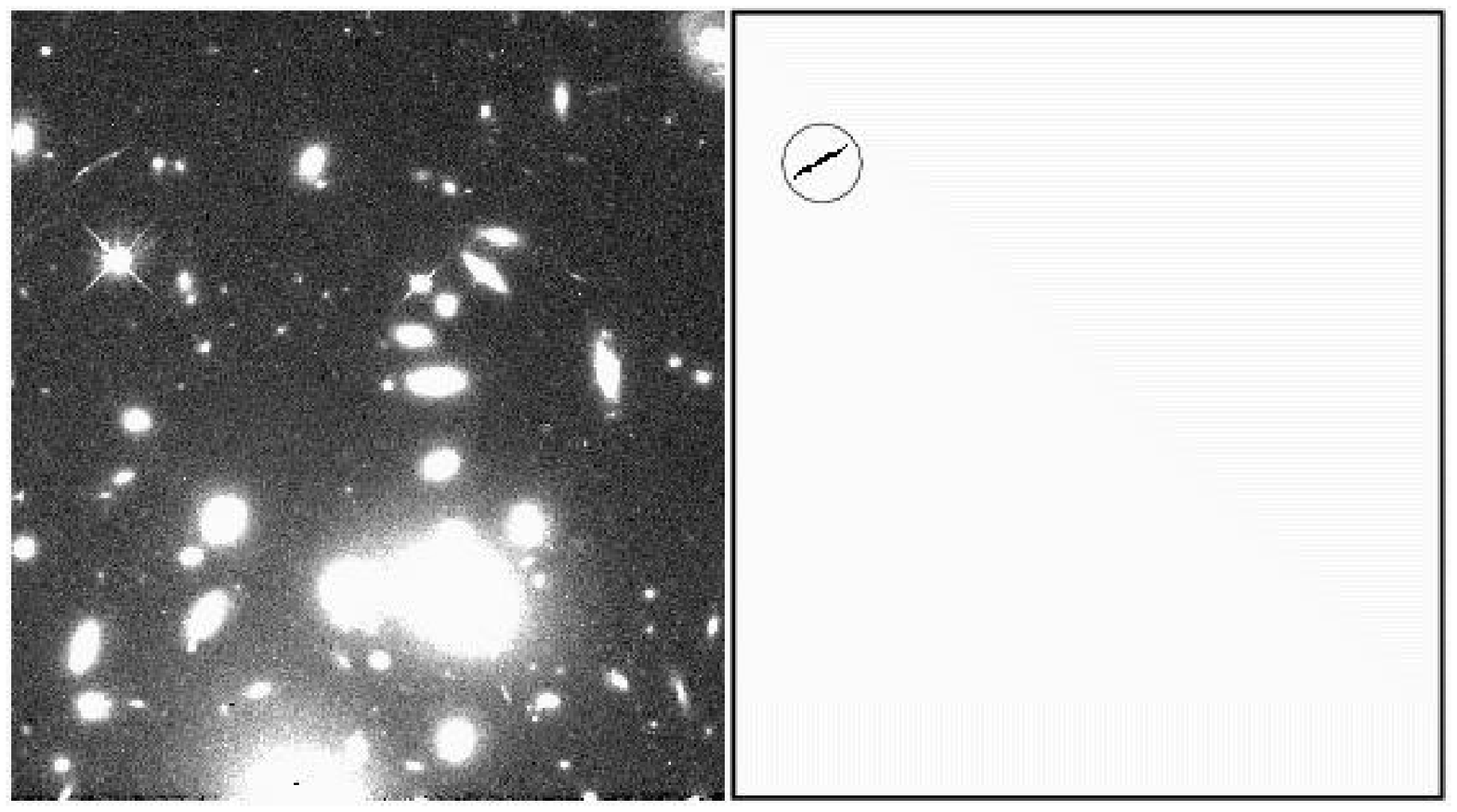}}
\subfigure[Abell 773]{
\includegraphics[width=6.5cm, angle=0]{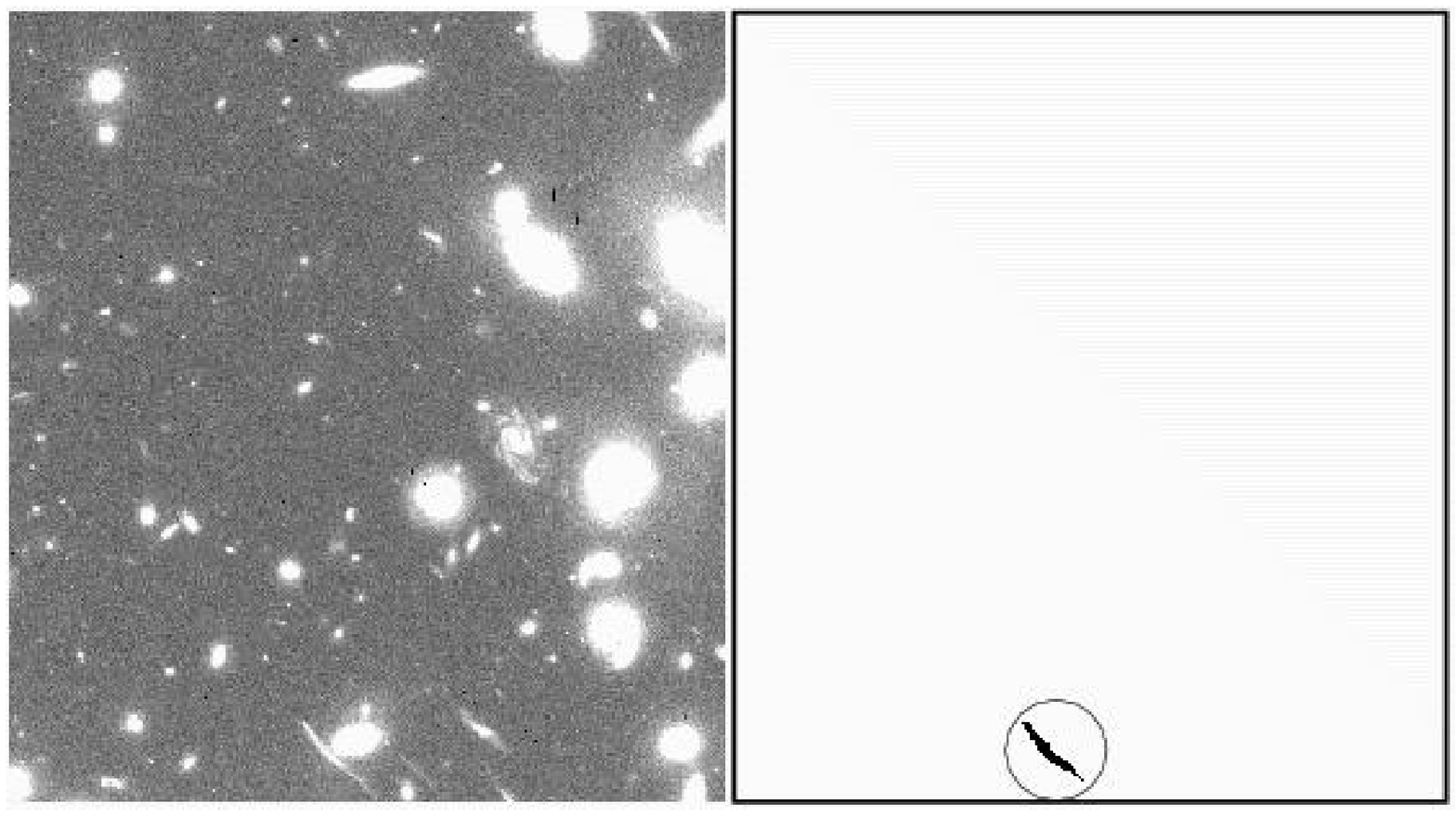}}
\hspace{0.5cm}
\subfigure[Abell 963]{
\includegraphics[width=6.5cm, angle=0]{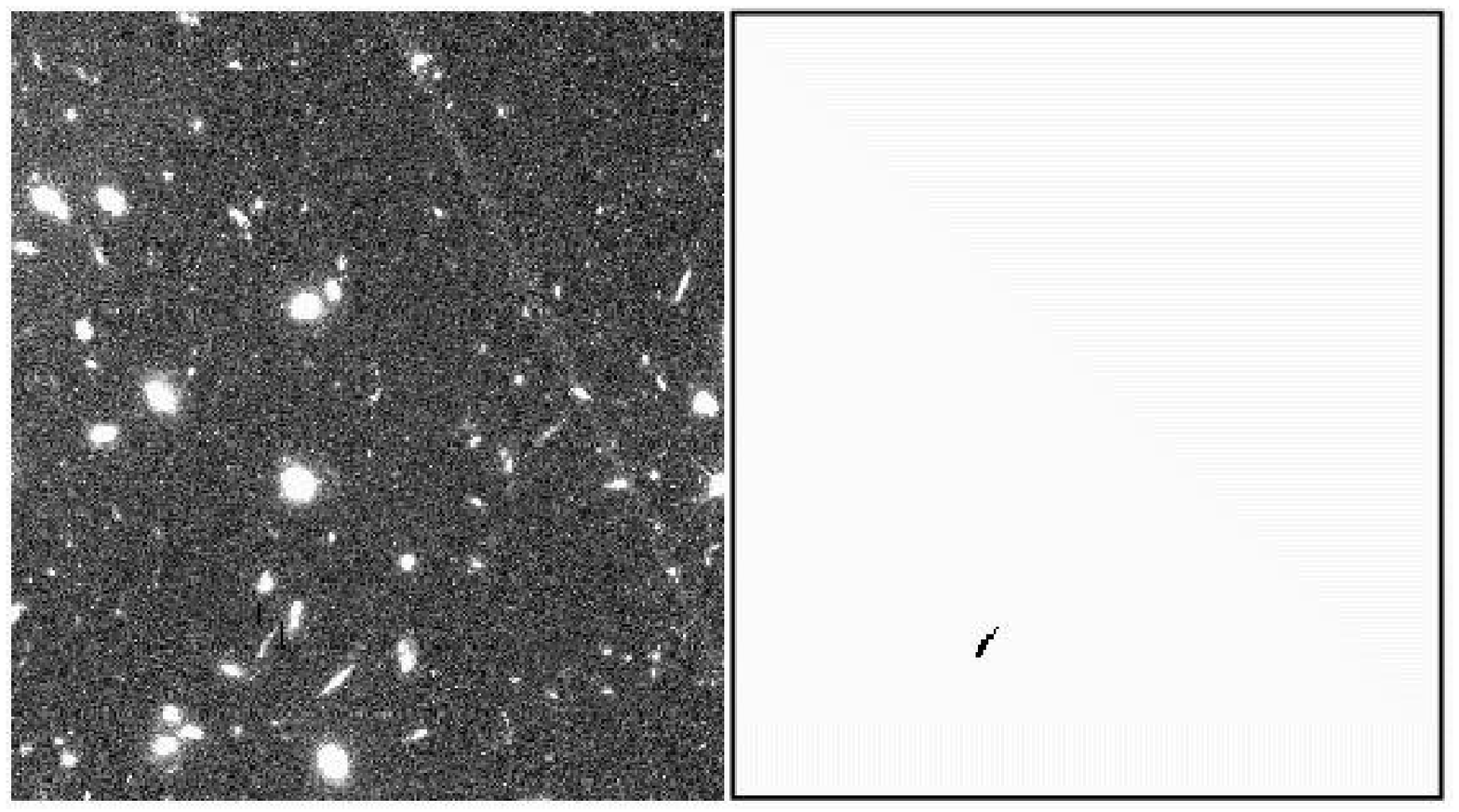}}
\subfigure[Abell 963]{
\includegraphics[width=6.5cm, angle=0]{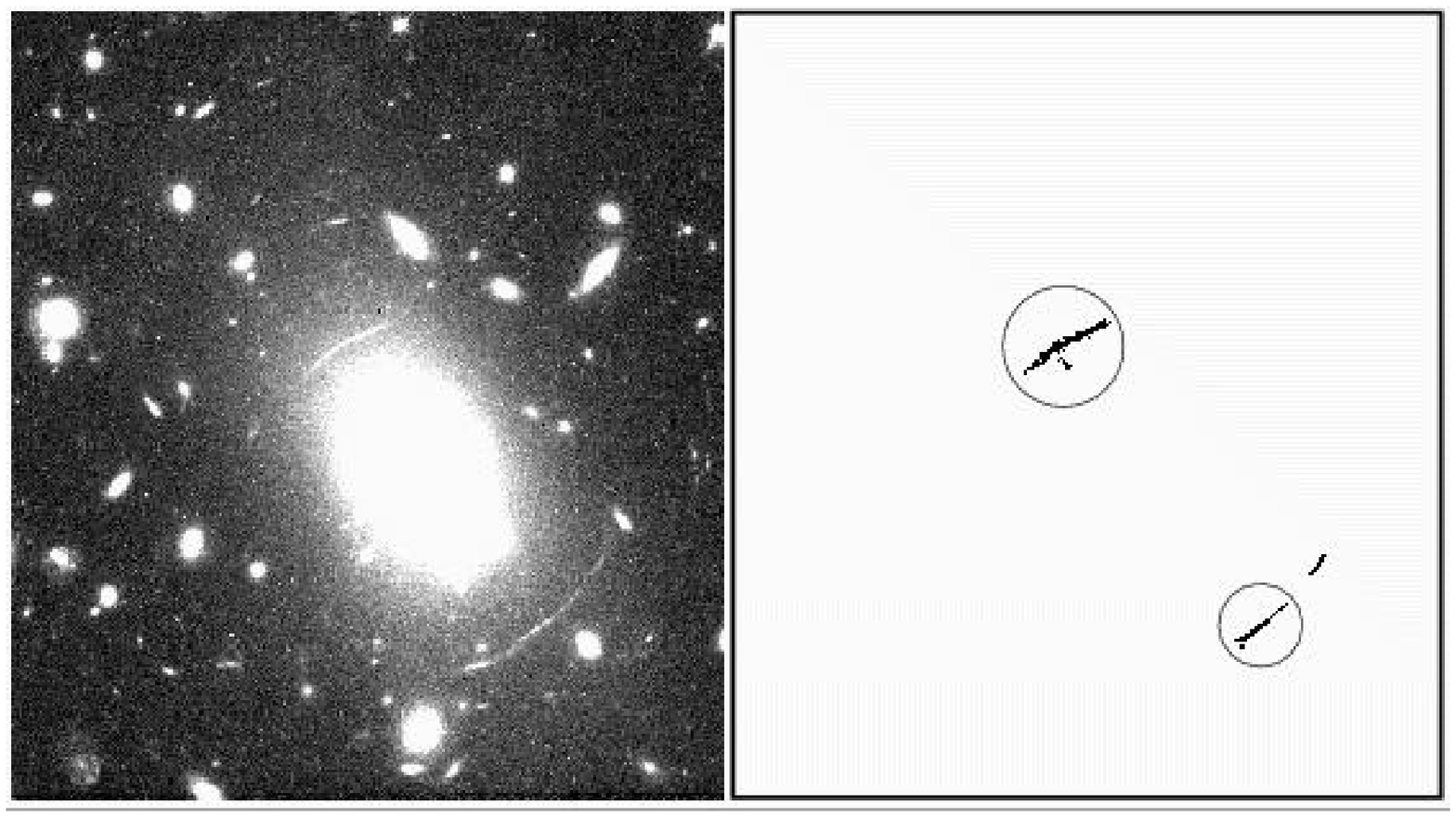}}
\hspace{0.5cm}
\subfigure[Abell 963]{
\includegraphics[width=6.5cm, angle=0]{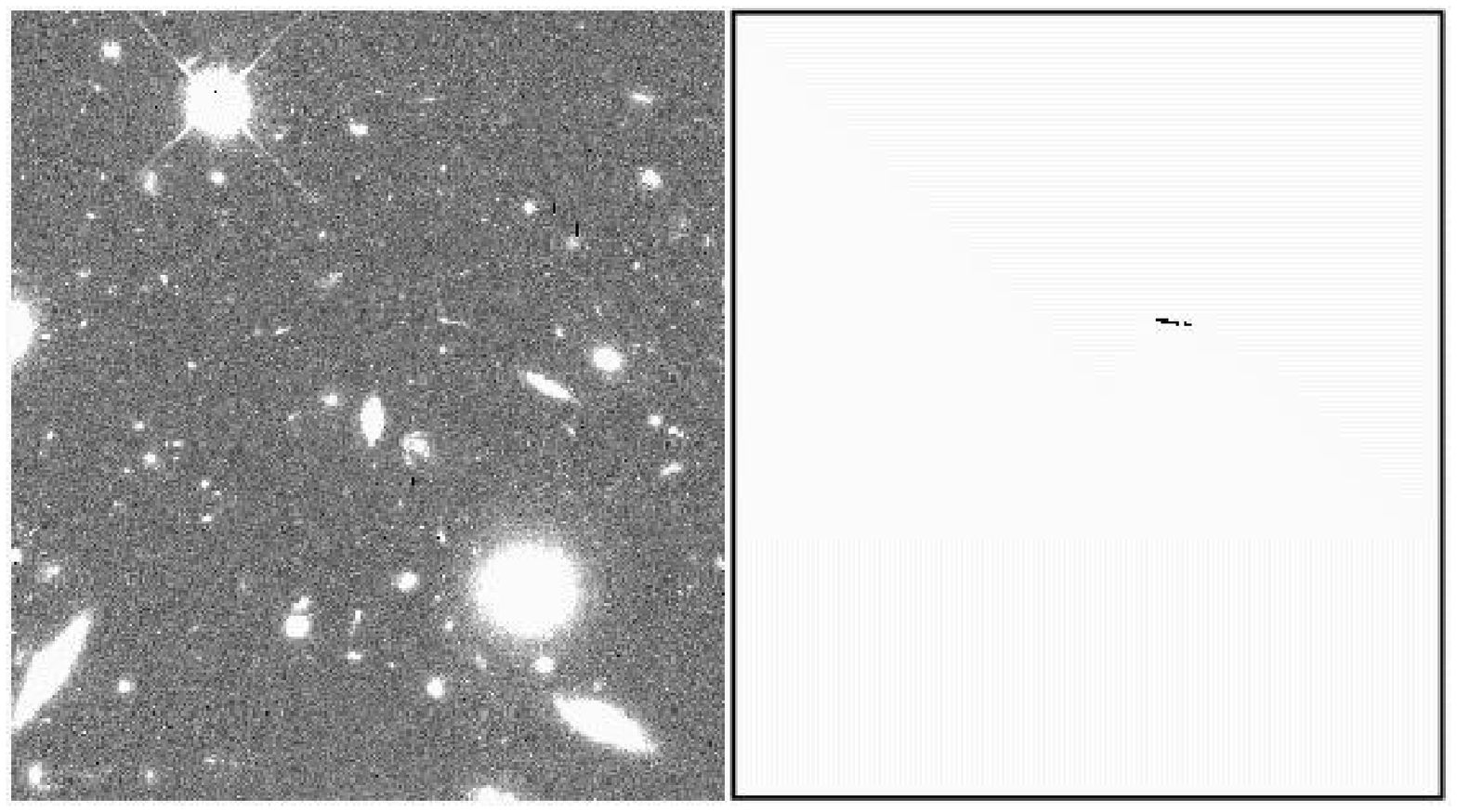}}
\subfigure[Abell 2218]{
\includegraphics[width=6.5cm, angle=0]{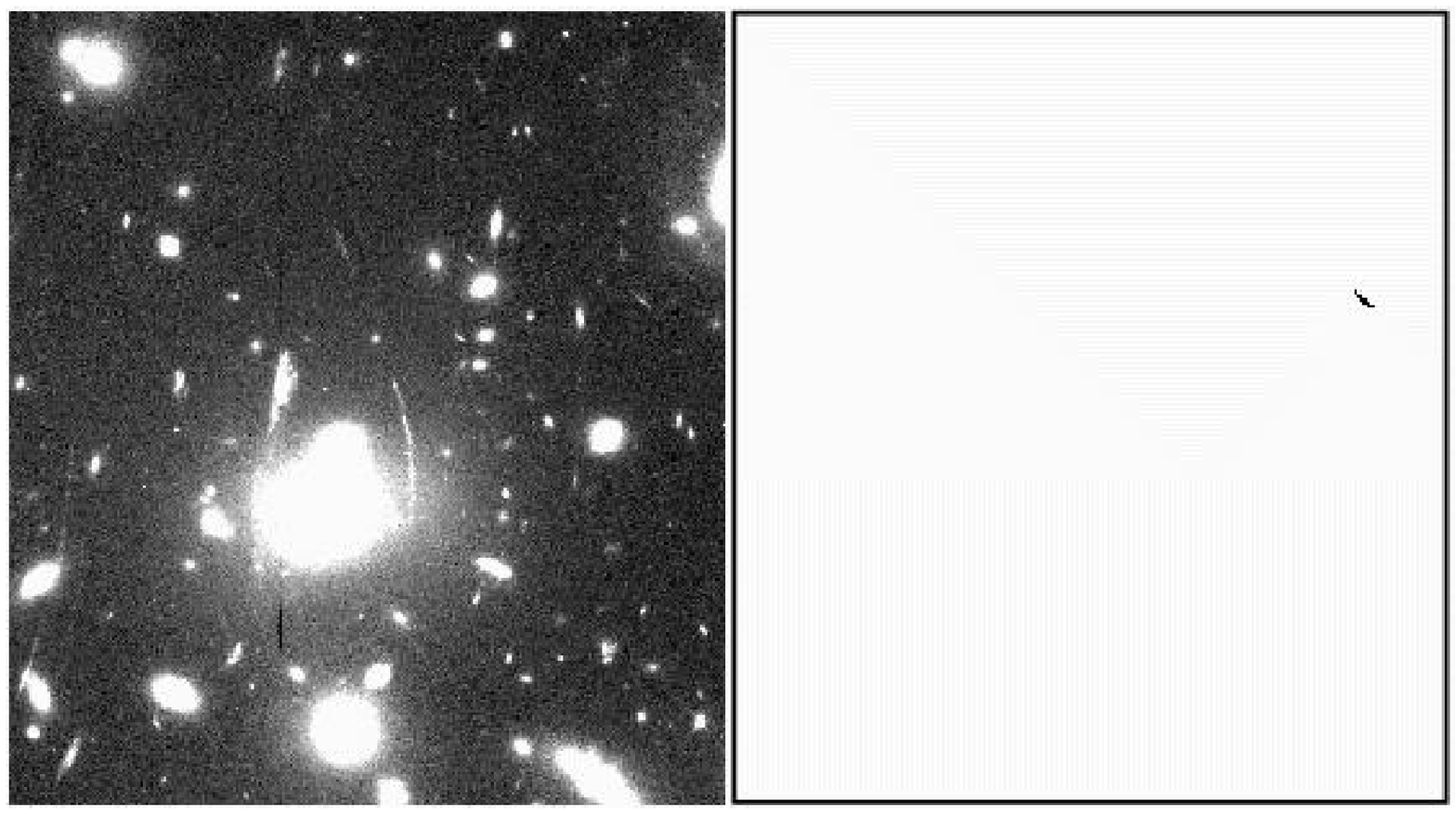}}
\hspace{0.5cm}
\subfigure[Abell 2218]{
\includegraphics[width=6.5cm, angle=0]{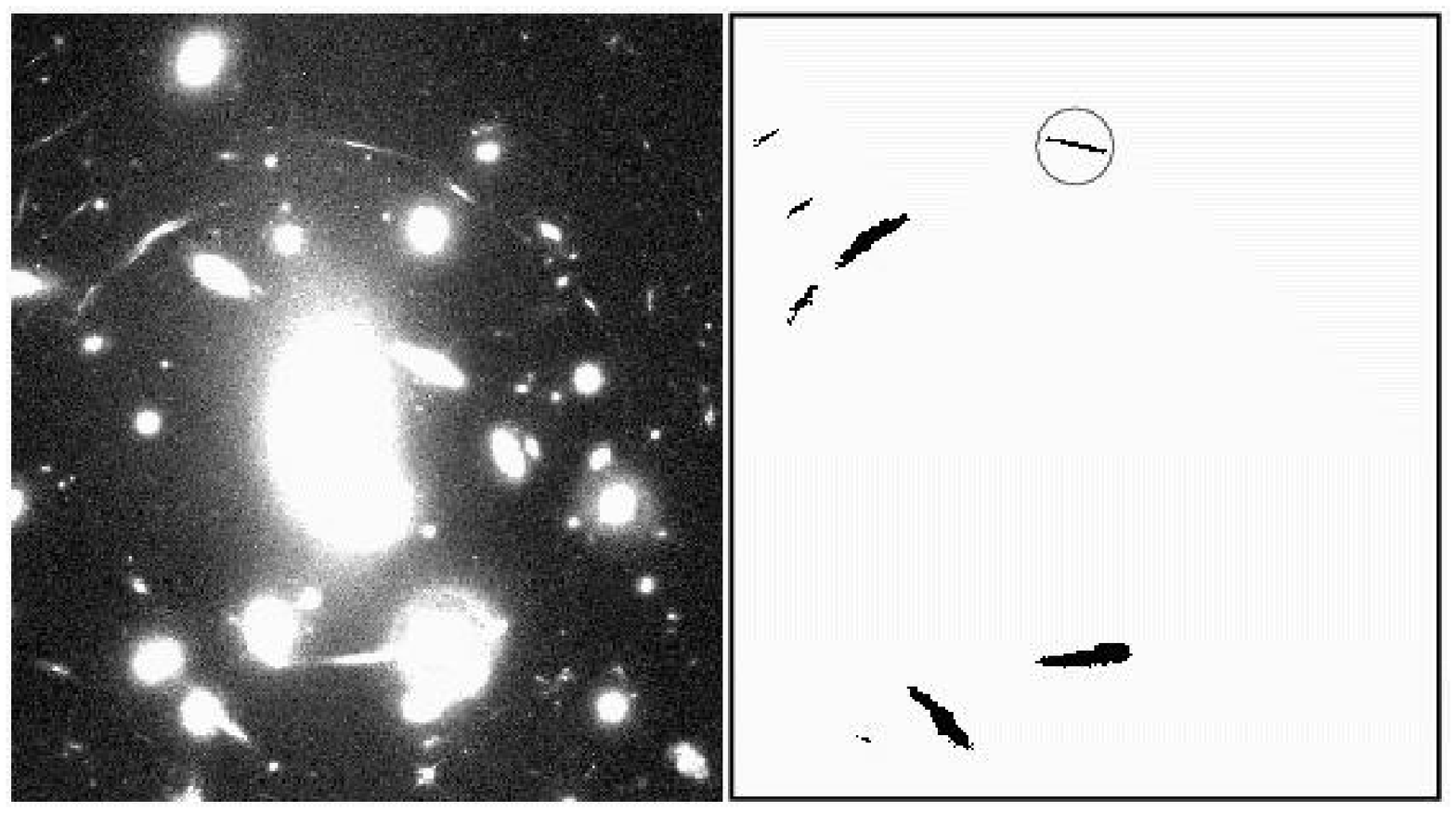}}
\subfigure[Abell 2218]{
\includegraphics[width=6.5cm, angle=0]{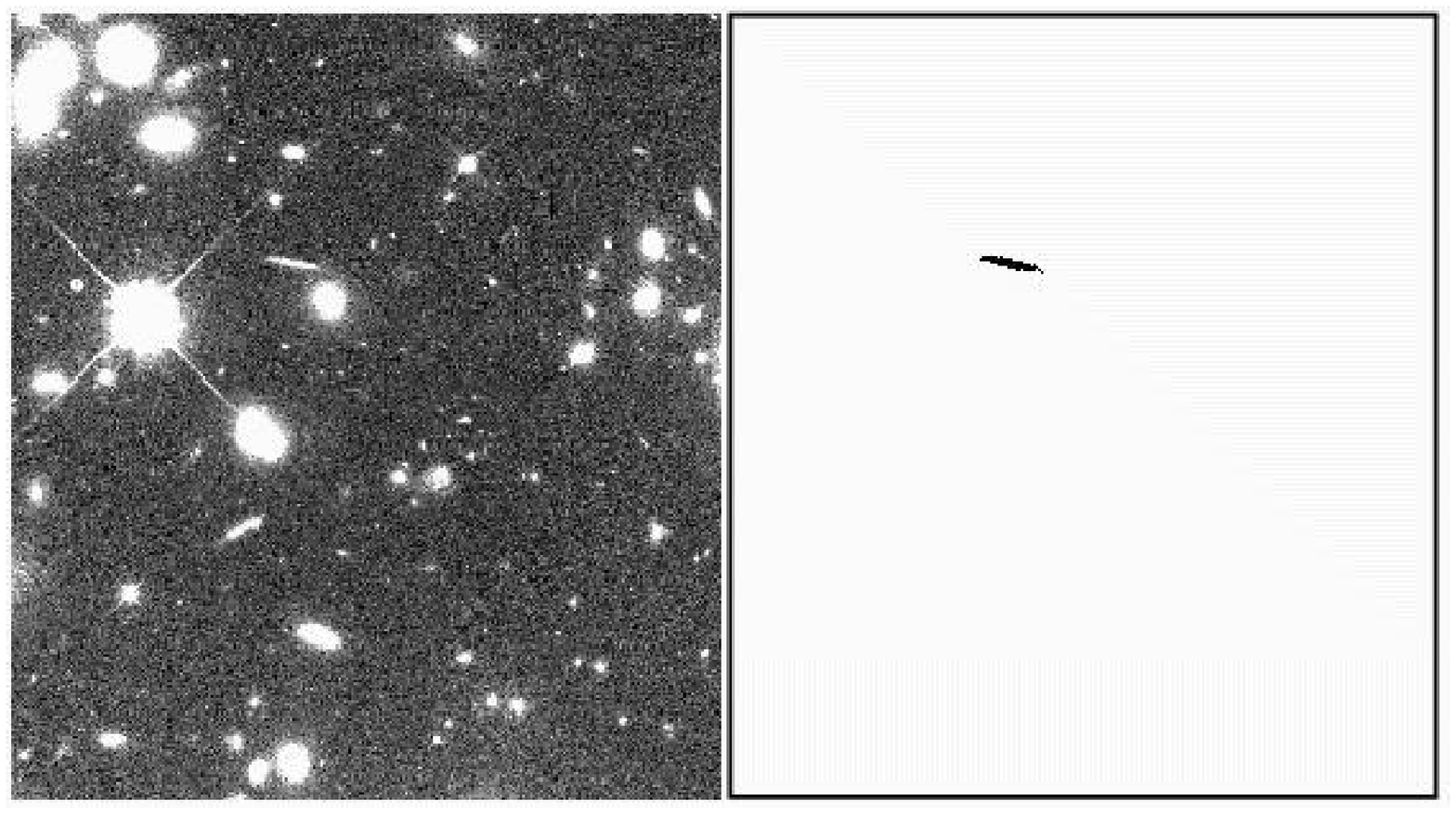}}
\caption{[continued]}
\end{figure*}

\begin{figure}[!ht]
\centering
\includegraphics[width=6cm, angle=-90]{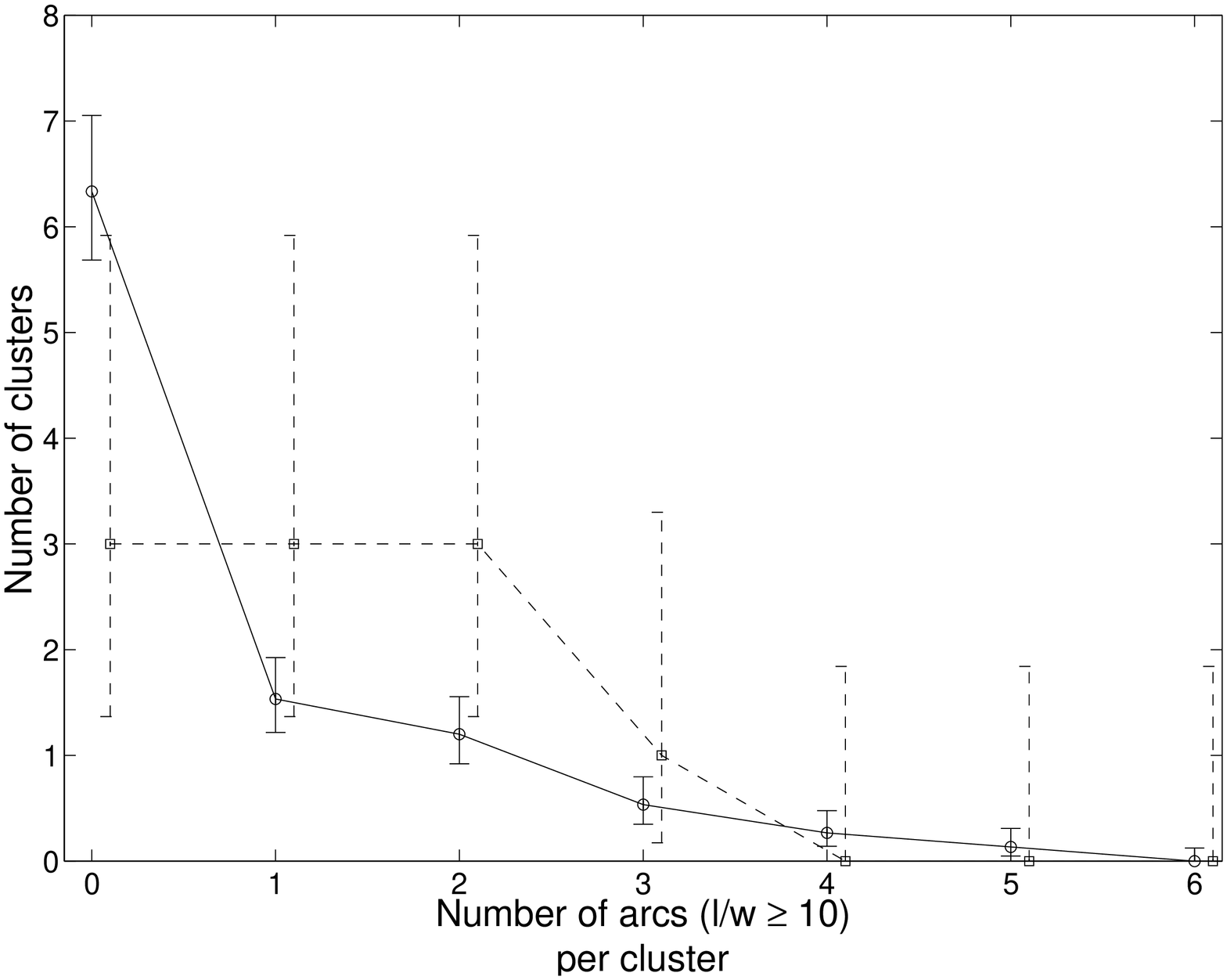}
\caption{Distributions of arcs per cluster with
  $l/w\geq 10$ for the observed (squares and dashed lines) and
  simulated (circles and solid lines) samples. The distribution for
  the simulated clusters has been normalized to the total number of
  observed clusters. Poisson errors are indicated.}
\end{figure}

Additional statistical properties are consistent between the observed
and simulated samples. Figure $7$ compares the distributions of arc
lengths, widths, and $l/w$ ratios. All of them show consistency
between the lensing efficiency predicted by the simulated $\Lambda{\rm
  CDM}$ clusters and the observed data.

\section{Conclusions}

We have carried out a comparison of the arc-production efficiency of
an observed sample of massive clusters, and of a sample of clusters
produced in a $\Lambda{\rm CDM}$ N-body simulation. The two samples are
roughly matched in mass. In our simulations we have used a realistic
source population, in the form of the HDF, including the number
density, surface brightness profiles, and redshifts of the galaxies in
it. We have attempted to simulate the observational
effects that may influence arc statistics. Finally, we have carried
out an objective comparison of the two samples, not limited by arc
brightness, by applying to both samples an automatic arc-finding
algorithm.

Our main result is that the lensing efficiencies of the observed and
the simulated clusters are consistent. The observed clusters may be
somewhat more efficient lenses than the simulated clusters based on
the total number of giant arcs they produce, or on the distribution of
the number of arcs per cluster, but the differences are marginally
significant ($1.6\sigma$ and $2.1\sigma$, respectively) and could well
be due to small number statistics. The above differences could also be
explained if the lensing efficiency correlates with mass, since the
observed clusters go to somewhat higher masses than the simulated
ones. We note, however, that the most massive cluster in the observed
sample, Abell 1835, has no detected giant arcs, and in the mass range
of our observed and simulated clusters the lensing efficiency may have
a weak dependence on cluster mass (e.g., Hennawi et al. 2005).

Given that we have used the same $z_{c}=0.2$ N-body cluster sample as
B98, it is instructive to compare our results. We find that the number
of giant luminous arcs produced by the clusters is higher by a factor
of $\sim 3$ than the number obtained by B98. The difference can be
explained by the different background source densities used by us and
by B98. According to the Fern{\' a}ndez-Soto et al. (1999) photometric
redshift catalog, the density of background sources with $R\leq 24$
mag and $0.6\leq z_{s}\leq 2$ is $6.5\times 10^{4} {\rm deg}^{-2}$,
higher by a factor of $3.2$ than the density assumed by B98. This
result, by itself, is a major step towards solving the
order-of-magnitude problem reported by B98. Other differences between
our work and B98 seem to have little or no influence on the number of
giant luminous arcs.  In particular, using a realistic redshift
distribution for the background sources, as opposed to placing them
all at redshift $z_{s}=1$, results in only a small change in the
number of giant luminous arcs. We should remember that, contrary to
other studies that have investigated the effect of source redshift
(e.g., Wambsganss et al. 2004, Dalal et al. 2004, Li et al. 2005), our
study also includes simultaneously other effects -- source size and
surface brightness -- which can also be redshift dependent, and which
can cancel or reinforce the influence of the redshift on cross
section. In any event, our simulations, which are more realistic in
this respect than previous ones, show that the total number of arcs is
weakly affected by placing all sources at $z_{s}=1 $ (or at
$z_{s}=1.5$), rather than at their true redshifts.

As opposed to previous studies, our approach has allowed us to compare
the observations and the theoretical predictions also for less
luminous arcs. As described above, for such arcs, which are several
times as numerous as the luminous ones, there is also agreement
between the observed and simulated samples, though with a hint for a
higher (factor $\lesssim 2$) lensing efficiency of the observed
sample.

Our results point the way to several improvements and extensions that
could be implemented in future studies, and which could provide
stronger tests of cosmology by means of arc statistics. From the
observational aspect, the uncertainties in the present work are
dominated by the small number of clusters and arcs in the observed
sample. Larger samples of clusters, observed to large depth and at
higher resolution with the HST Advanced Camera for Surveys, are
becoming available, and could be used for this purpose. Furthermore,
the effects of cluster selection criteria on arc statistics need to be
investigated. For example, it could be that X-ray luminosity selects
clusters that are undergoing merging and that such clusters have
enhanced lensing cross-sections. This could then explain the factor
$\sim 2$ enhancement in lensing efficiency of observed clusters
suggested by our results, if this enhancement is confirmed. Arc
statistics measured in cluster samples selected by other means (e.g.,
by ``red sequence'' colors; Gladders \& Yee 2000) could test this
hypothesis.

On the theoretical side, N-body simulations with higher temporal and
spatial resolution can give a more reliable picture of cluster
substructure and of the importance of merging events for lensing
efficiency. A larger number of clusters and of projections of each
clusters are needed, in view of the large scatter we have found in the
number of arcs produced by different clusters and by different
projections of the same clusters. The mass of the cD galaxy, should
also be included in the lensing simulations (Meneghetti et al. 2003),
as should the cluster gas component (Puchwein et al. 2005).

Although the background source population we have used is more
realistic then those used by previous studies, this can still be
improved. The HDF covers a small area (which also forced us to tile it
in our simulations), and as a results it may be unrepresentative of
the typical source density or redshift distribution. Furthermore,
several more magnitudes of depth would allow probing the source
population of lensed arcs more reliably. The Hubble Ultra Deep Field
could serve for this purpose, once its photometric redshifts have been
measured and calibrated.

In summary, we have shown that model $\Lambda{\rm CDM}$ clusters at
$z_{c}=0.2$ have lensing efficiencies similar as to those of observed
X-ray selected clusters of similar mass and redshift. Improved arc
statistics tests can be based on larger, differently selected,
observational samples, and on larger and more sophisticated
simulations. Our results do not address the separate issue of whether
N-body simulations produce the correct number and mass distribution of
clusters as a function of redshift. Although this aspect of the
problem was also folded into previous analyses, it would best be
examined by means of a direct comparison of the models and the
observations, independent of lensing properties.
\begin{figure*}[!ht]
\centering
\subfigure{
\includegraphics[width=6cm, angle=-90]{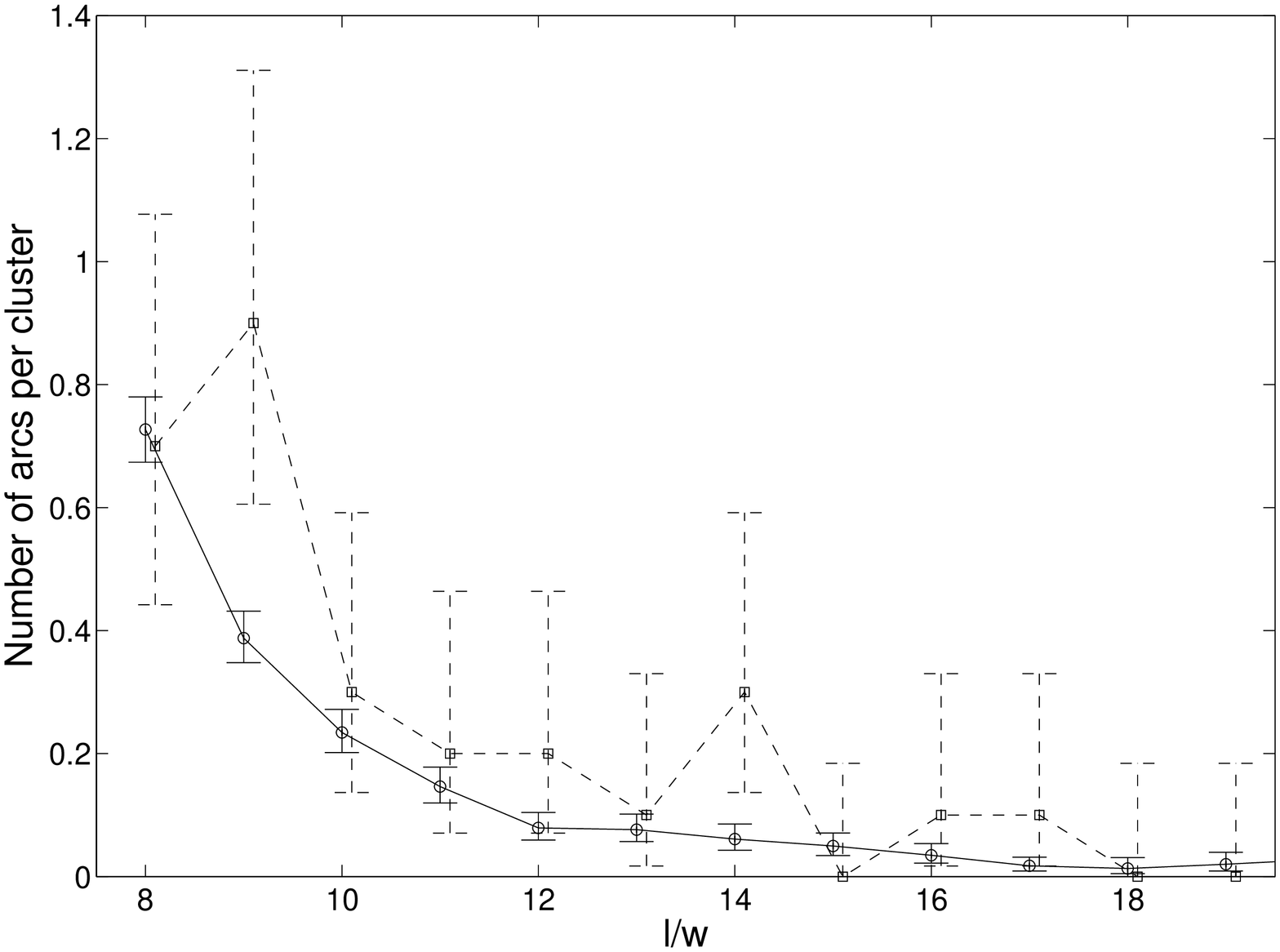}}
\hspace{1cm}
\subfigure{
\includegraphics[width=6cm, angle=-90]{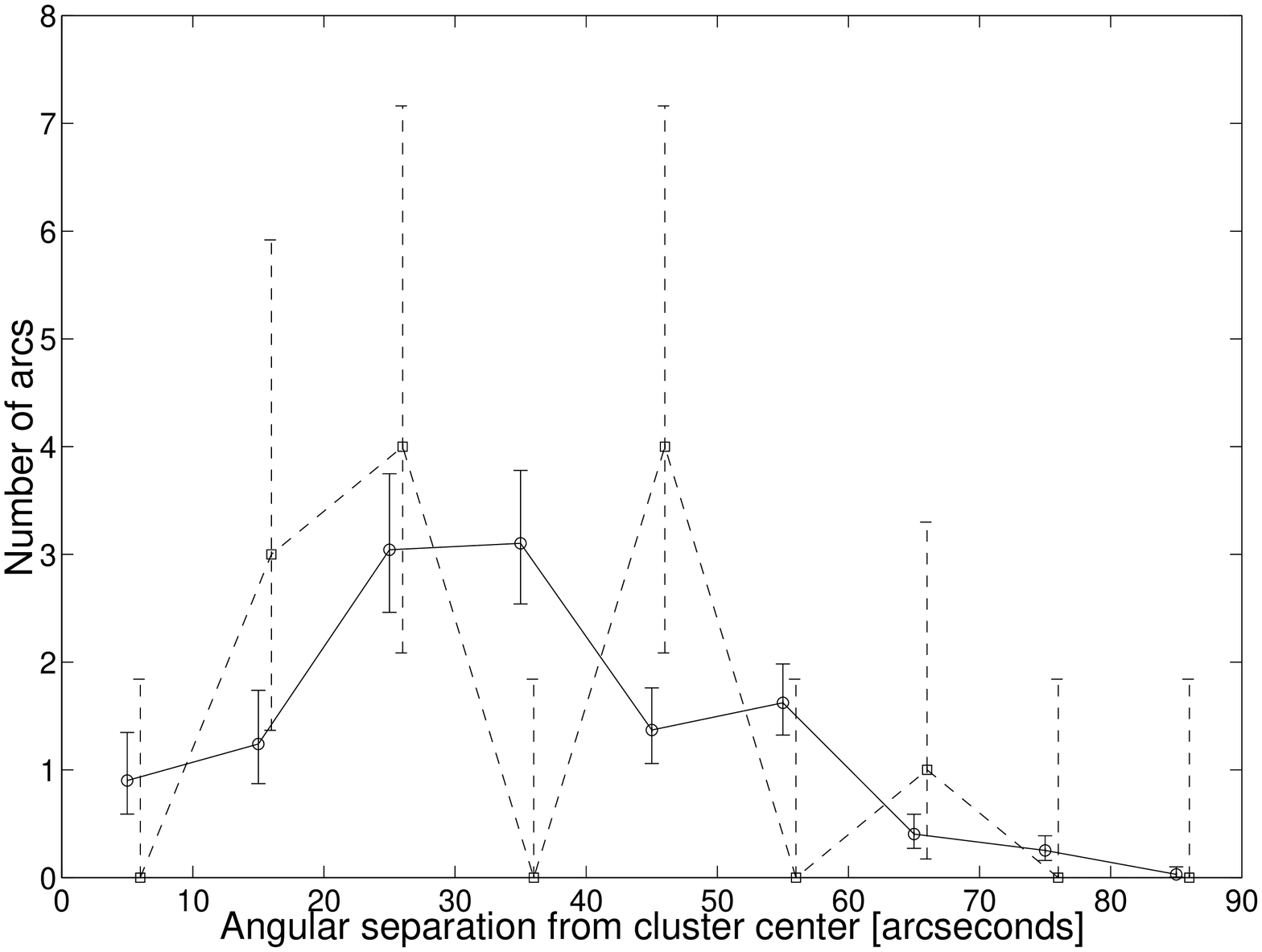}}
\subfigure{
\includegraphics[width=6cm, angle=-90]{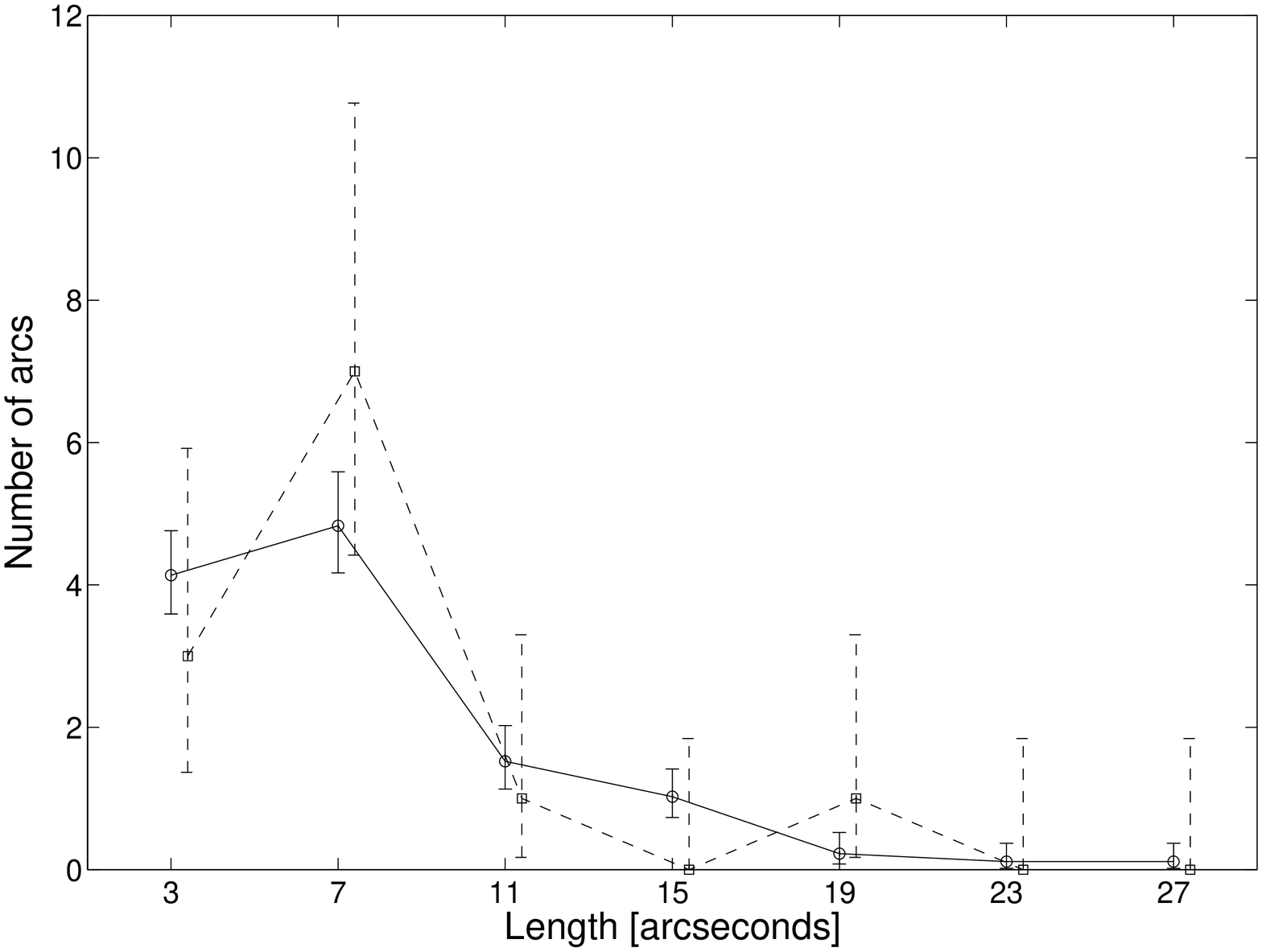}}
\hspace{1cm}
\subfigure{
\includegraphics[width=6cm, angle=-90]{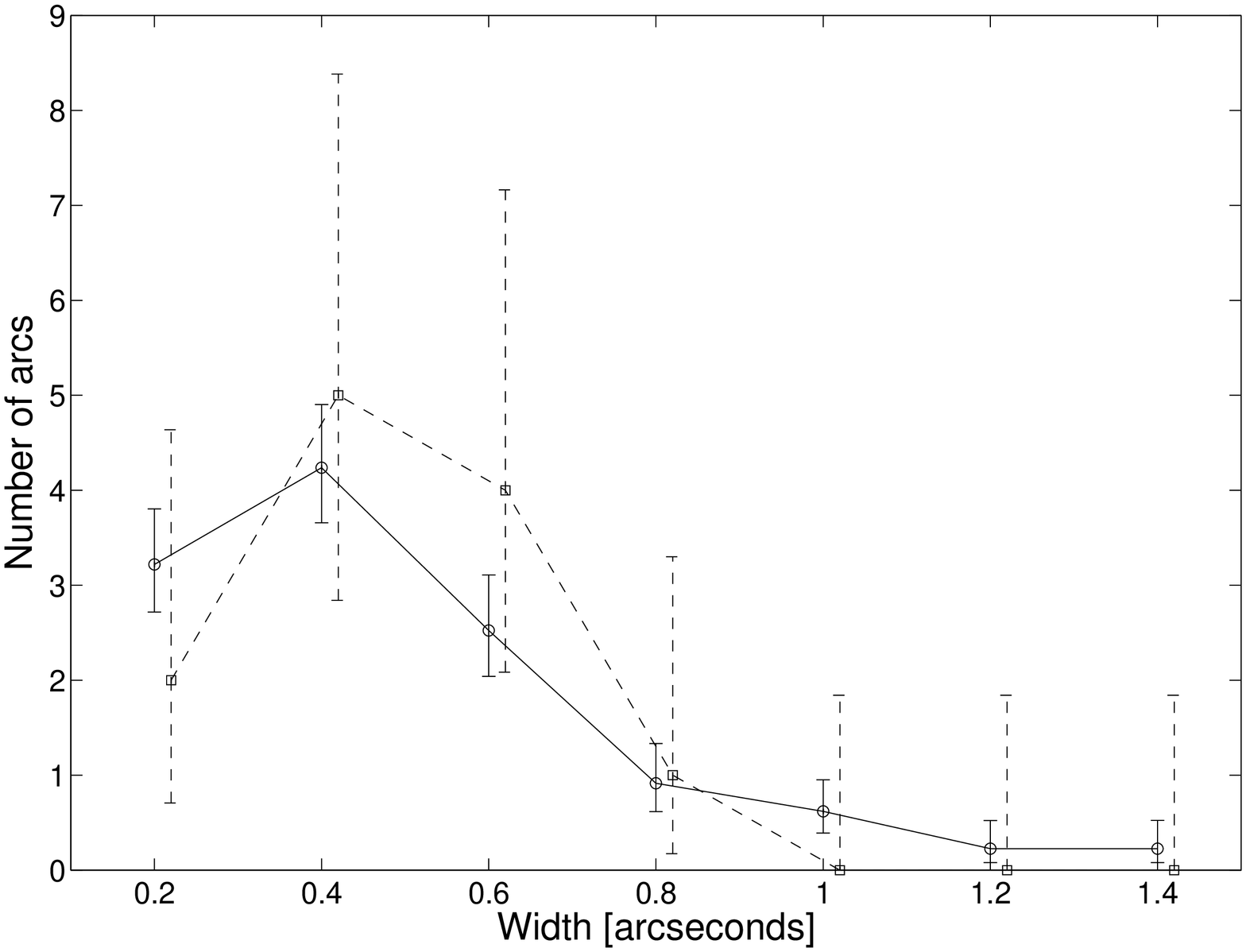}}

\caption{Distributions of arc properties. The distributions of lengths, widths, and
  angular separations from cluster center, are for giant arcs
  ($l/w\geq 10$). The solid lines represent the simulated sample and the
  dashed lines the observed sample.}
\end{figure*}

\section*{Acknowledgements}

We thank A. Gal-Yam and the anonymous referee for comments. This work
was supported by the German Israeli Foundation for Scientific Research
and Development.

\appendix
\section{The arc-finding program}

In order to allow an objective and quantitative comparison of arc
statistics in our simulations and the observed dataset, and to permit
arc searches in large real and simulated datasets, we have devised an
automatic arc-finding algorithm. As mentioned in $\S4$, our algorithm
is implemented in a simple script that makes repeated calls to
SExtractor (Bertin $\&$ Arnouts 1996). The script is written in the C
language and utilizes the CFITSIO software
library\footnote{http://heasarc.gsfc.nasa.gov/fitsio} (Pence, 1999).
The program is available at http://wise-obs.tau.ac.il/~assafh, or on
request from the authors. The outline of our algorithm is as follows.

Our algorithm is based on a series of six consecutive detection and
elimination SExtractor calls, each using slightly different
parameters, as listed in Table A1. After each call, the detected
objects which have axis ratios smaller than a set threshold, as listed
in Table A1, are eliminated from the detection image. Calls 1, 3, and
5, are applied to the original image, while calls 2, 4, and 6, are
applied to the output of calls 1, 3, and 5, in which eliminated
objects have been replaced by a constant background. We then combine
the three images produced by calls 2, 4, and 6, into a segmentation
image in which all non-zero pixels are assigned a constant value. We
calculate the $l/w$ ratio, as described in $\S4$, of the objects
detected by SExtractor in the combined image.  Objects below a
user-defined $l/w$ threshold are eliminated from the detection image,
thus producing a final image of detected arcs.
\begin{table}[!ht]
\caption{Sextractor parameters sets used in the detection and
  elimination processes}
\smallskip
\begin{center}
\begin{tabular}{ccccccc}
\hline
\noalign{\smallskip}

 & \multicolumn{6}{c}{Detection and elimination calls}\\

\noalign{\smallskip}

Parameter name & 1 & 2 & 3 & 4 & 5 & 6 \\
\noalign{\smallskip}
\hline
\noalign{\smallskip}
DETECT\_MINAREA & 35 & 35 & 20 & 20 & 20 & 20\\
DETECT\_THRESH & 2$\sigma$ & 2$\sigma$ & 3$\sigma$ & 3$\sigma$
& 4$\sigma$ & 4$\sigma$ \\
FILTER & Yes & No & Yes & No & Yes & Yes \\
DEBLEND\_NTHRESH & 6 & 2 & 6 & 2 & 24 & 24 \\
DEBLEND\_MINCONT & 0.003 & 0.01 & 0.003 & 0.01 & 0.001 &
0.001 \\
\noalign{\smallskip}
\hline
Axis ratios eliminated after detection & $<2$ & $<3.3$ & $<2$ & $<3.3$ & $<2$ & $<3.3$ \\
\noalign{\smallskip}
\hline
\end{tabular} 
\end{center}

\end{table}

A different arc finding algorithm has recently been devised by Lenzen
et al. (2004). In order to evaluate and compare the performance of our
algorithm and of the one of Lenzen et al. (2004), we have applied both
of them to the HST images of Abell 2218 and Abell 1689. We set the
detection threshold of our algorithm to $l/w\geq 5$ and accordingly
set the eccentricity detection parameter to $0.8$ in the Lenzen et al.
arc detector, while the rest of their arc detection parameters were
set according to Fig.  15 from Lenzen et al. (2004). Both algorithms
required approximately the same computation time. For example
detecting arcs in the $1430\times 1430$ pixel images of Abell 2218,
required $\sim 1$ min of CPU time, running on a computer with an Intel
Pentium 4, 2.4 GHz, processor, and with 1 GB of RAM. There is good
agreement between the arcs detected by both algorithms. However, the
properties of the detected arcs are different. The Lenzen et al.
algorithm tends to ignore low surface brightness regions in arcs.
Thus, the areas of the arcs detected by the Lenzen et al. arc finder
are usually smaller than of those detected by our algorithm. As a
result, the arc magnitudes and geometrical properties, such as $l/w$,
will be different. In addition, the arcs detected by the Lenzen et al.
arc finder tend to have higher eccentricities, resulting sometimes in
false detection of spurious artifacts (see Fig. 15 of Lenzen et al.
2004). These ``false positives'' are non-arc objects with
intrinsically low eccentricity, but for which the algorithm has
isolated a high-eccentricity core. In Fig. A1, we illustrate the
performance of both algorithms on the two different clusters. For the
HST image of Abell 1689, also used for illustration by Lenzen et al.,
close arcs and arcs which lie next to a galaxy are better resolved by
the Lenzen et al. algorithm, while our arc finder sometimes tends to
detect two arcs as one, and to associate part of a nearby galaxy with
an arc. These differences in performance are similar to the
differences already pointed out by Lenzen et al.  themselves, when
comparing their algorithm to a direct application of SExtractor to
cluster images.  However when applied to the data for Abell 2218, it
appears that the Lenzen et al. program picks out only the highest
surface-brightness areas in the arcs, while our arc finder performs
better.
\begin{figure}[!ht]
\centering
\subfigure[Abell 1689]{
\includegraphics[width=14cm]{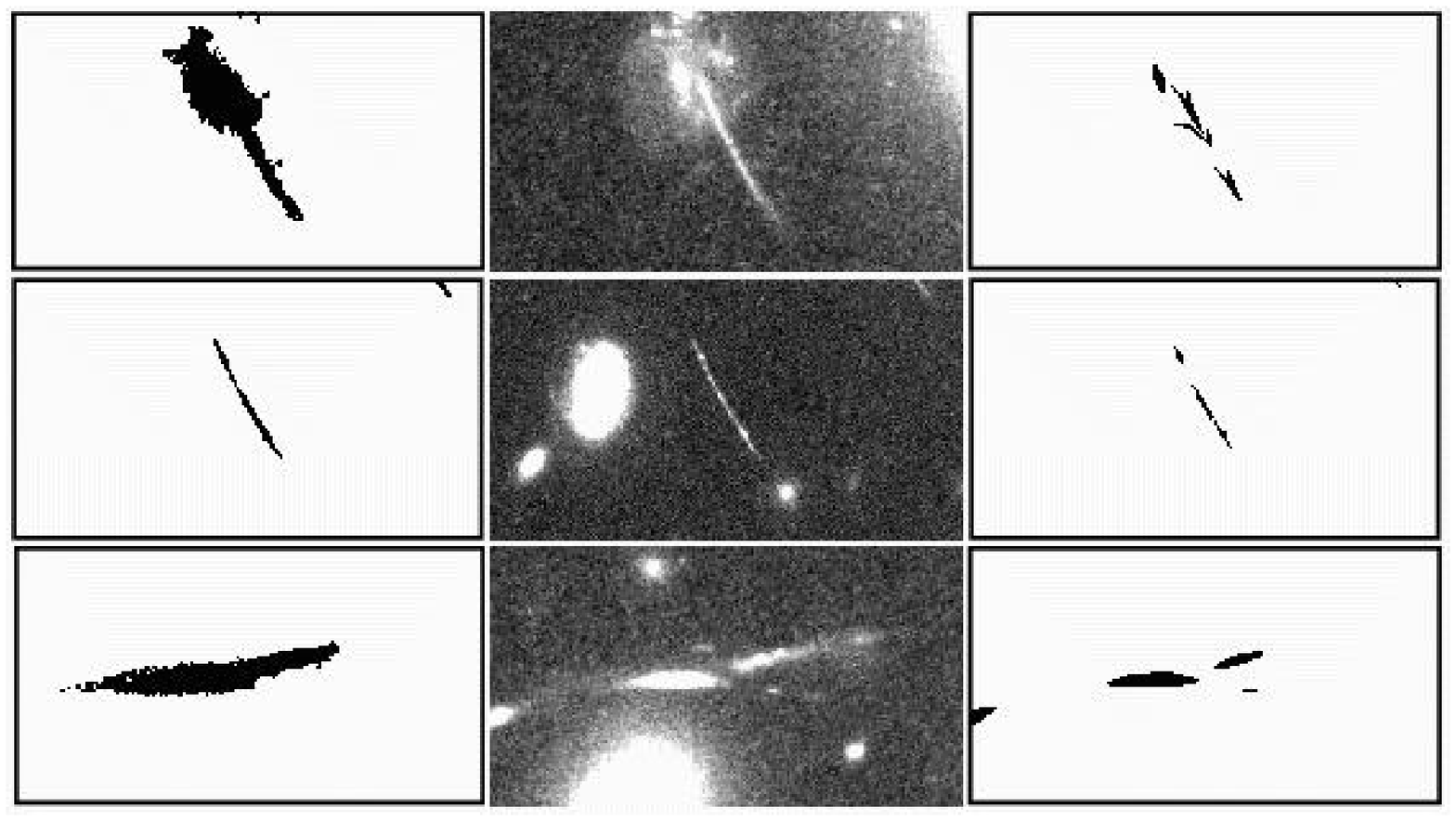}}
\subfigure[Abell 2218]{
\includegraphics[width=14cm]{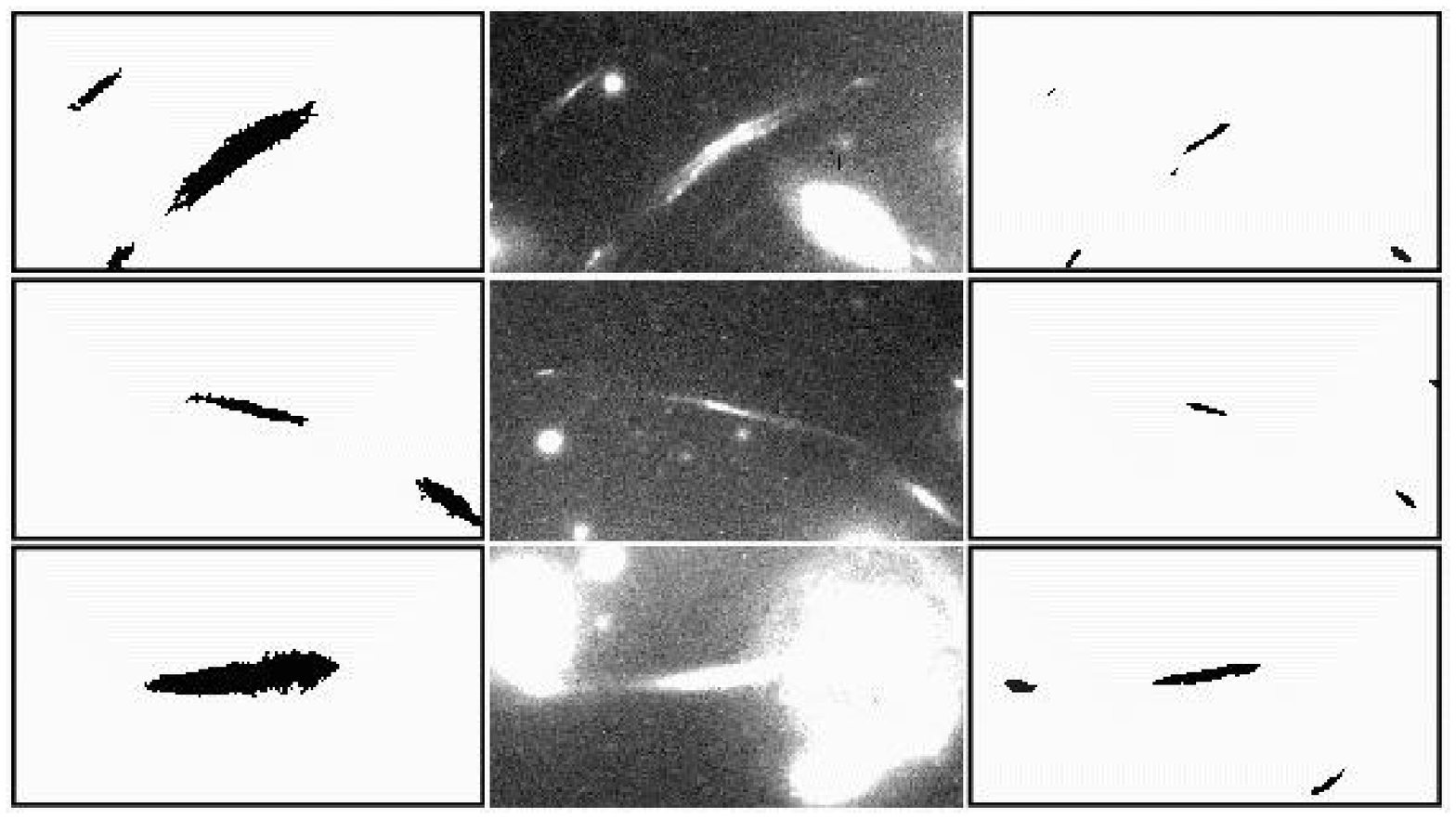}}
\caption{Comparison between arc detection results obtained with our
  arc finder (left panels) and with the Lenzen et al. (2004) arc
  detector (right panels). Although generally the same arcs are
  identified by both algorithms, our arc finder identifies more of the
  surface area associated with each arc.}
\end{figure}

To obtain an estimate of the detection efficiencies of both
algorithms, we applied them, with a detection threshold of $l/w\geq
10$, to $10$ lensing realizations of one of the projections of the
simulated cluster cj1409. The ``real'' arcs were identified by
applying our algorithm to each image before the addition of
observational effects. We then compared the detection results of both
algorithms by comparing them to the ``real'' arcs by eye.  Whenever a
``real'' arc or a part of it was detected, it was assigned a detection
weight of $1$, even in cases when it was split into two arcs in the
images with the observational effects. Fig. A2 shows the fraction of
arcs detected by each algorithm, as a function of magnitude. As seen
in Fig. A2, the detection efficiencies of the two algorithms are
similar.  However, as in the application to the real cluster data, the
arcs in the detection images produced by the Lenzen et al. algorithm
constitute a small fraction of the total arc areas. In the balance,
while the Lenzen et al. arc finder sometimes achieves superior
resolution, our algorithm identifies arcs with areas more similar to
those picked out by human eyes, and produces fewer false detections.
\begin{figure}[!ht]
\centering
\includegraphics[width=8cm, angle=-90]{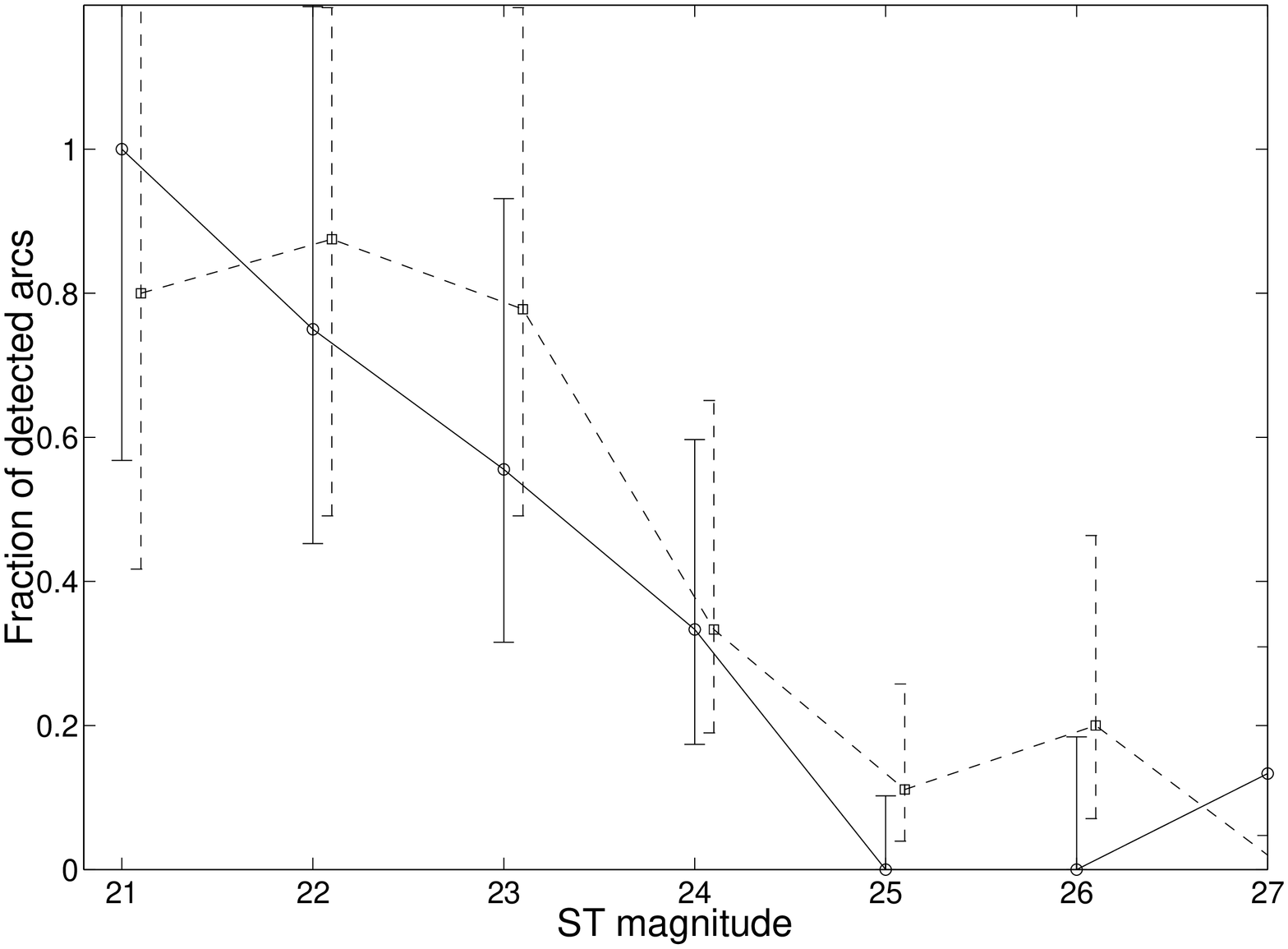}
\caption{Giant arcs $(l/w \geq 10)$ detected in $10$
  simulated lensed images with observational effects as a fraction of
  the arcs found in the same images without observational effects, for
  our arc-finding algorithm (solid line), and the Lenzen et al. (2004)
  algorithm (dashed line). Poisson errors are indicated.}
\end{figure}


\end{document}